\documentclass{article}
\pdfoutput=1
\usepackage{graphicx}
\begin{document}

\def\kms{\mathrm {km\, s}^{-1}}
\def\msun{\rm M_{\odot}}
\def\bh{{\rm BH}}
\def\bhp{{\rm BH,1}}
\def\bhs{{\rm BH,2}}
\def\bht{{\rm BH,T}}
\def\gw{{\rm GW}}
\def\df{{\rm DF}}
\def\inf{{\rm inf}}
\def\hard{{\rm hard}}
\def\mbh{M_{\rm BH}}
\def\etal{{et al.}}
\def\simlt{\mathrel{\rlap{\lower 3pt\hbox{$\sim$}}\raise 2.0pt\hbox{$<$}}}
\def\simgt{\mathrel{\rlap{\lower 3pt\hbox{$\sim$}} \raise 2.0pt\hbox{$>$}}}
\def\lsim{\mathrel{\rlap{\lower 3pt\hbox{$\sim$}}\raise 2.0pt\hbox{$<$}}}
\def\gsim{\mathrel{\rlap{\lower 3pt\hbox{$\sim$}} \raise 2.0pt\hbox{$>$}}}
\def\di{\mbox{d}}
\def\mbulge{M_{\rm Bulge}}
\def\msunpc3{\msun~{\rm {pc^{-3}}}}
\newcommand{\be}{\begin{equation}}
\newcommand{\ee}{\end{equation}}

\title{{\Large \bf Massive Binary Black Holes in the Cosmic
Landscape}}

\date{    }
\maketitle
\begin{center}
\vskip -1.5truecm

{Monica Colpi$^{1}$ \& Massimo Dotti $^{2}$}\\
\vskip 0.3truecm

{\small 1-Department of Physics G. Occhialini, University of Milano-Bicocca,
\break Piazza della Scienza 3, 20126 Milano, Italy, colpi@mib.infn.it\\
2-Department of Astronomy, University of Michigan, \break Ann Arbor, MI, 48109, USA, mdotti@umich.edu\\}
\end{center}

\begin{abstract}
{\it Binary} black holes occupy a special place in our quest for
understanding the evolution of galaxies along cosmic history.  If
massive black holes grow at the center of (pre-)galactic structures
that experience a sequence of merger episodes, then dual black holes
form as inescapable outcome of galaxy assembly, and can
in principle be detected as powerful dual quasars.  
But, if the black holes reach {\it
coalescence}, during their inspiral inside the galaxy remnant,
then they become the loudest sources of gravitational waves ever in the
universe.  The {\it Laser Interferometer Space Antenna} is being
developed to reveal these waves that carry information on the mass and
spin of these binary black holes out to very large
look-back times.  Nature seems to provide a pathway
for the formation of these exotic binaries, and a number of key
questions need to be addressed: How do massive black holes pair in a
merger? Depending on the properties of the underlying galaxies, do
black holes always form a close Keplerian binary?  If a binary forms,
does hardening proceed down to the domain controlled by gravitational
wave back reaction? What is the role played by gas and/or stars in
braking the black holes, and on which timescale does coalescence
occur? Can the black holes accrete on flight and shine during their
pathway to coalescence?  After outlining key observational facts on
dual/binary black holes, we review the progress made in tracing their
dynamics in the habitat of a gas-rich merger down to the smallest
scales ever probed with the help of powerful numerical simulations.
N-Body/hydrodynamical codes have proven to be vital tools
for studying their evolution, and progress in this field is 
expected to grow rapidly in the effort to describe, in full realism, the physics 
of stars and gas around the black holes, starting from the cosmological 
large scale of a 
merger.
If detected in the new window provided by the upcoming gravitational
wave experiments, binary black holes will provide a deep view into the
process of hierarchical clustering which is at the heart of the
current paradigm of galaxy formation. They will also be exquisite
probes for testing General Relativity, as the theory of gravity.  The
waveforms emitted during the inspiral, coalescence and ring-down phase
carry in their shape the sign of a dynamically evolving space-time and
the proof of the existence of an horizon.

\end{abstract}


\section{Black holes in the cosmic landscape}

\subsection{Introduction}

Massive black holes weighing from a million to a billion suns have
long been suspected to be the powerhouse of energetic phenomena taking
place in distant galaxies. Energy in the form of radiation, high
velocity plasma outflows, and ultra relativistic jets, is extracted
efficiently from the gravitational field of the black hole when gas is
accreting from the parsec scale of a galactic nucleus down to the
scale of the horizon.  Since the early discovery of quasars, the
accretion paradigm has been at the heart of the interpretation of
massive black holes as being "real" sources in the universe.  It was
also recognized in the late 60's that the luminous quasars and the
active galactic nuclei (AGNs) were undergoing strong cosmic evolution:
nuclear activity was common in the past and declined with cosmic
time. No bright quasar lives in our local universe, but a few AGNs 
are present with very
modest activity, representing the fading tail of the population.  From
simple considerations on the life-cycle of quasars, there has been
always the suspicion that at high redshifts accretion was ignited in
many if not all galaxies, leading to the commonly held premise that
most galaxies we see in our local universe should host in their
nucleus a massive black hole, relic of an earlier active phase.

\bigskip
For long times, black hole masses in AGNs and quasars have been
inferred only indirectly using as chief argument the concept of
Eddington limited accretion.  But today, due largely to the impact of
ground-based telescopes and of the Hubble Space Telescope, the
mass of quiescent black holes inside the cores of nearby galaxies
including our own Milky Way, has been measured using stars and/or gas
clouds as dynamical probes.  Now there is indeed strong circumstantial
evidence that massive black holes are ubiquitous in ellipticals and in
the bulges of disk galaxies.  Astronomers discovered in addition, and
most importantly, the existence of tight correlations between the
black hole mass and the large scale properties of the underlying
host$^{1-7}$. 
It is currently believed that during the formation of galaxies, a
universal mechanism was at play able to deposit, in the nuclear
regions, large amounts of gas to fuel the black hole to such an extent
that its feedback, i.e. its large-scale energy/momentum injection,  had blown the gas
out, thus sterilizing 
the galaxy against further star
formation$^{8-10}$.
Major galaxy mergers could be at the heart of this
symbiotic relationship as they may explain both
the ignition of a powerful AGN and the formation of a violently
perturbed galaxy remnant dominated by stellar dispersion$^{11}$.  

\bigskip

Galaxy formation is a genuine cosmological problem: the cooling and
collapse of baryons in dark matter halos, clustering hierarchically,
is the prime element for understanding galaxy evolution.
The time of first appearance of black holes in mini-halos is largely
unknown: whether they formed at redshift $z\sim 20$ as relic of the
first metal free stars$^{12}$, or later in more massive virialized haloes
from unstable gaseous disks or dense young star clusters, is
unknown$^{12-20}$. What is currently known is that black holes 
mainly grow from gas accretion$^{21,22}$, and that bright quasars, hosting a
billion solar mass black hole, are already in place out to redshift
$z\sim 6$ when the universe was $\lsim 10^9$ years old$^{23}$.  The new
paradigm of the concordant  evolution of black holes and galaxies 
imposes a new perspective: black holes previously believed to
play a passive role are now in "action" shaping their cosmic
environment$^{24-25}$.

\bigskip
The coevolution of black hole and galaxies embraces a variety of
astrophysical phenomena that are now becoming of major scientific
interest. They go from the formation of black hole seeds in the first
pre-galactic structures clustering hierarchically at very high
redshifts, to black hole growth and feedback in major gas-rich
mergers.  But not only. A new and critical aspect of this concordant
evolution is the presence of {\it black hole pairs} in galaxies that
form during the violence of a merger.  {\it There is growing evidence
that Nature provides through mergers the privileged habitat where
massive binary black holes can form and evolve. } But why are {\it
binary black holes} important ?  The answer is manyfold and is the focus 
of this writing. 

\bigskip
The review is organized as follows. In Section 1.2 we introduce key
physical scales of black hole binaries on their path to coalescence
under the emission of gravitational waves. In Section 1.3 we summarize
current evidence of dual and binary black holes in the realm of
observations.  Section 1.4 reports on the quest for the presence of
massive black hole binaries in bright elliptical galaxies. Section 2
describes the basic physics of black hole inspiral both in gas-poor
and gas-rich environments also with use of numerical
simulations. Section 3 summarizes selected results attained over the
last years in the study of black hole hardening in pure
stellar/collisionless backgrounds.  Section 4 shows key results on the
formation of black hole pairs/binaries during gas-rich major as well
as minor mergers, simulated with unprecedented accuracy starting from
cosmologically motivated initial conditions.  In Section 5 we address
a number of issues related to the dynamics of binary black holes in
massive gaseous disks: orbital decay, eccentricity evolution,
accretion on flight and migration in a circum-binary disk.  Section 6
explores the delicate and important issue on the electromagnetic
counterparts of gravitational wave coalescence events. Section 7
contains our conclusions and future perspectives.

\subsection {Gravitational waves from binary black holes}

Einstein's theory of space-time and gravity, General Relativity,
predicts that motions of masses produce propagating waves that travel
through space-time at the speed of light $c$.  Two masses (of total
mass $M_\bht$) in a binary emit gravitational waves that in the far
field zone, i.e. much farther away from the source, perturb space-time 
with a dimensionless metric-strain amplitude $h_\gw \sim
(M_\bht V_{\rm cir}^2)/D,$ where $D$ is the distance and $V_{\rm cir}$ the orbital velocity 
(in geometric units).  This indicates that the strongest sources of gravitational
waves are the most massive bodies that move close to the speed of
light, i.e.  the most massive black holes ($M_\bht\sim
10^6-10^9\,\msun$) that move in a binary, with $V_{\rm cir}\sim c,$ at
the moment of their coalescence.  Therefore, binary black holes
"created" in merging galaxies offer one of the most promising venues
of producing the loudest gravitational waves in the universe that the
{\it Laser Interferometer Space Antenna} ({\it LISA}) or/and the Pulsar
Timing Array (PTA) experiment are expected to detect $^{26-34}$.  During
late inspiral and final plung-in, the black holes become genuine
dynamical entities, creating rich space-time structure that remains
imprinted in the strain amplitude signal.  The detection of this
signal gives unique probe of the existence of the horizon which
clothes the interior singularity of the black hole.

\bigskip

According to the {\it no-hair} theorem, a black hole is similar to an
elementary particle specified uniquely by its mass $M_\bh$ and
angular momentum ${\bf J}_{\bh}$ (dimensionless spin ${\hat
a}=cJ_\bh / GM^2_{\bh}\leq 1$ ). The size of the horizon that varies
from $2GM_\bh/c^2$ for ${\hat a}=0$ to $GM_\bh/c^2$ for ${\hat a}=1$
(for a maximally rotating Kerr hole) is very tiny compared to the galactic
scale: for ${\hat a}=0$, the horizon is $3\times 10^{11}
(M_\bh/10^6\,\msun)\,{\rm cm}$, i.e.  only $10^{-7}(M_\bh/10^6\,\msun
)\,{\rm pc}$ (considering that one pc is the mean distance between
the stars in the solar neighborhood).  Thus only through the detection
of the waveforms emitted from "coalescing" and "vibrating" black holes
we may hope to unravel their true nature$^{35}$.  With {\it LISA}, black hole
masses and spins will be measured with astonishing accuracy$^{36,37}$, and
this claim suffices to explain why binary black holes are becoming now
so important not only for exploiting the black hole demography in a
cosmological context, but gravity in the strong field regime.  As two
black holes of mass $M_{\bh,1}$ and $M_{\bh,2}$ coalesce, a new black
hole forms with mass less than the sum $M_{\bh,1}+M_{\bh,2}$ according
to the area theorem.  Gravitational waves carry away mass-energy and
angular momentum, and depending on the spin, they carry away also
linear momentum as the resulting radiation pattern is anisotropic. The
binary center of mass thus recoils back and the relic black hole
acquires a kick$^{38-43}$. While for slowly spinning black holes and for spins
(anti-)aligned with the orbital axis, recoil velocities are $\lsim
400\,\kms,$ there are spin configurations for which the recoil can be
as large as 4000 $\kms$, i.e in excess of the escape velocity from
the largest galaxies: this occurs when the black holes have comparable
masses, and the spin vectors have opposite directions and lie in the
orbital plane.

\bigskip

{\it Coalescence} and {\it recoil} are the final step of a long
process of {\it slow inspiral}.  According to the quadrupole formula,
a binary of total mass $M_\bht$, mass ratio $q_\bh=M_\bhs/M_\bhp\leq
1,$ semi-major axis $a,$ and eccentricity $e,$ is expected to coalesce,
i.e. to reduce its axis $a$ virtually to zero, after a time span$^{44}$
\begin{equation}
t_{\rm GW}={5\over 256 f(e)}{c^5\over G^3}{(1+q_\bh)^2\over q_\bh}{ a^4
\over M^3_\bht},
\end{equation}
where $f(e)=[1+(73/24)e^2+(37/96)e^4](1-e^2)^{-7/2}$.  For two
black holes of $M_\bht=10^6\,\msun$ released during a galaxy merger on
a circular orbit at a distance $a\sim 10$ pc, this time far exceeds
the age of the universe ($t_\gw\sim 10^{25}$ yr).  It is thus clear
that coalescing binary black holes are interesting astrophysical
sources only if nature provides a dissipative mechanism that guides
the inspiral from the scale of a merger, of $\sim 100$ kpc, to the
scale $a_\gw$ below which gravitational waves are conducive to
coalescence on a timescale $t_{\rm GW}\lsim 10^{10}$ yr: eq. (1)
yields
\begin{equation}
a_\gw= 2\times 10^{-3} f(e)^{1/4}{q^{1/4}_\bh\over(1+q_\bh)^{1/2}}\left ({M_\bht \over 10^6\,\msun}\right )^{3/4}\left( {t_\gw \over 10^{10}{\rm yr}}\right)^{1/4}\,\,{\rm pc} 
\end{equation}
corresponding to $2\times 10^{4}$ Schwarzschild radii for a black
hole of mass $M_\bht=10^6\,\msun$.  At $a_\gw$, the orbital period 
(for $q_\bh=1$ and $e=0$) is
of only
\begin{equation}
P(a_\gw)=2\pi\left( {a_\gw^3\over GM_\bht}\right )^{1/2}\sim  5 
\left ({M_\bht\over 10^6\,\msun}\right )^{5/8}\left( {t_\gw \over 10^{10}{\rm yr}}\right)^{3/4}\,\,{\rm yr}
\end{equation}
and the relative circular velocity 
\begin{equation}
V_{\rm cir}=\left ({GM_\bht\over a_\gw}\right )^{1/2}\sim  1742
\left ({M_\bht\over 10^6\,\msun}\right )^{1/8}\left( {t_\gw\over 10^{10}{\rm yr}}\right)^{-1/8}\,\kms 
\end{equation}
is far in excess of the dispersion velocity of the ambient stars in a
typical galaxy ($\sim  100-300\,\kms$).  The dynamic range for
describing the hole's inspiral is so overwhelmingly large and extreme,
going from the kpc scale down to $a_\gw$, that a variety of
processes may conspire to bring (or halt) the black holes down to
(or away from) coalescence.  Thus, the main focus of this review is to
highlight the {\it physical} and {\it dynamical} processes that guide
the sinking of massive black holes during binary mergers down to
$a_\gw$.

\subsection{Dual, binary and recoiling AGNs in the realm of observations}


\begin{figure}
\centering{
\includegraphics[width=0.5\textwidth]{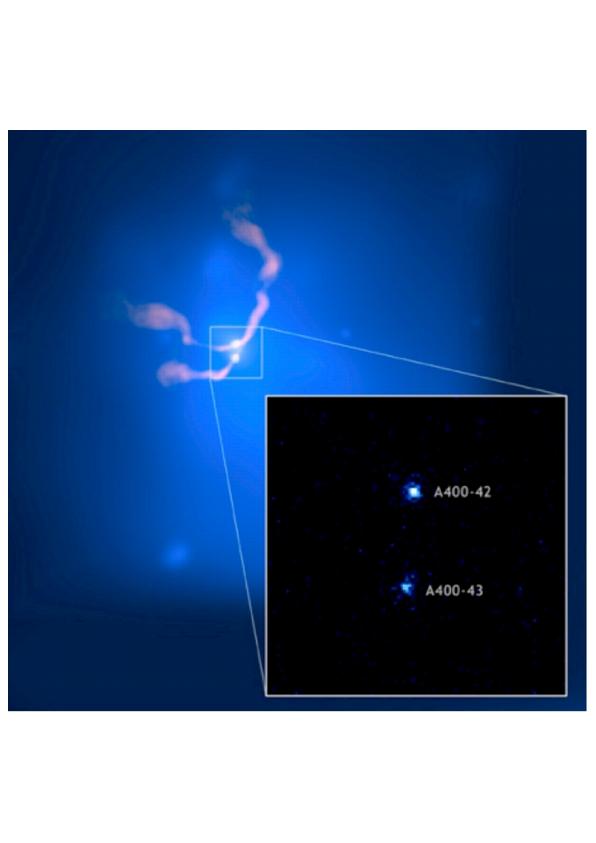}
\resizebox{6.5cm}{!}{\includegraphics{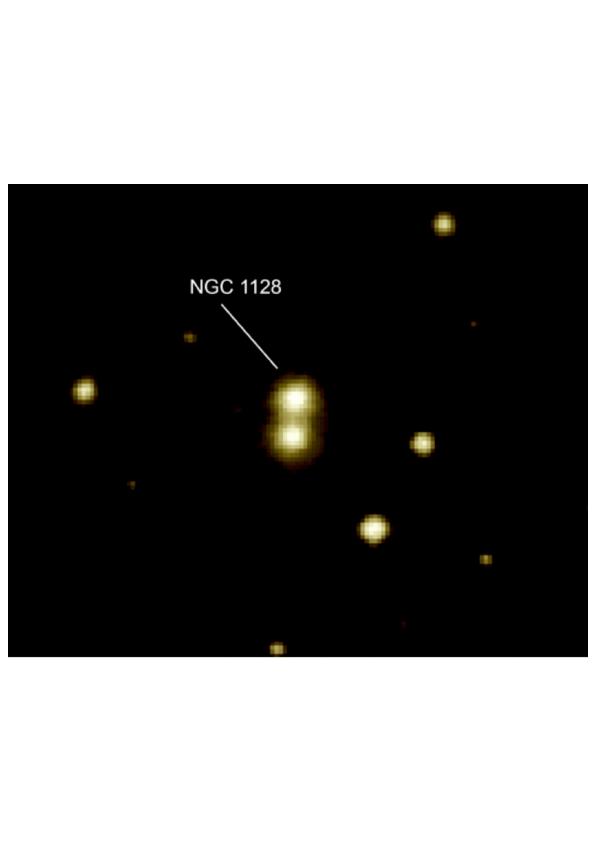}}\\
\caption{Composite X-ray (blue)/radio (pink) image of the galaxy
cluster Abell 400 
shows radio jets immersed in a vast cloud of
multimillion degree X-ray emitting gas that pervades the cluster. The
jets emanate from the vicinity of two supermassive black holes (bright
spots in the X-ray enlarged image). These black holes are in the
dumbbell galaxy NGC 1128 (whose optical image is on the right), which
has produced the giant radio source, 3C 75.}  }
\label{fig:3c75}
\end{figure}


\begin{figure}
\centering{
\resizebox{6cm}{!}{\includegraphics{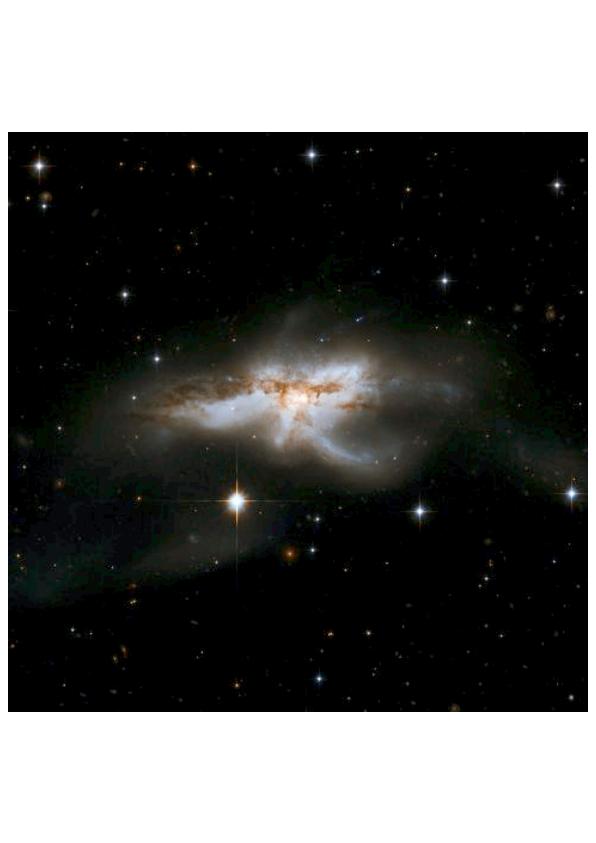}}
\resizebox{6cm}{!}{\includegraphics{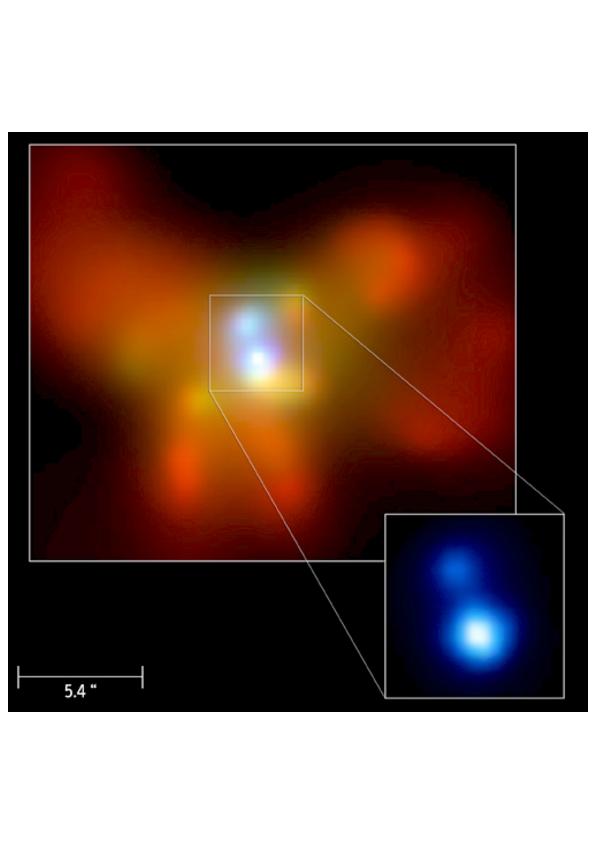}}\\
\caption{Optical image from {\it HST}
(left).  X-ray image$^{58}$ from {\it Chandra} (right) of the central region
of NGC 6240 showing the presence of two active nuclei: red color
refers to soft (0.5-1.5 keV), green to medium (1.5-5 keV), and blue to
hard (5-8 keV) X-rays. 1" corresponds to 700 pc in the galaxy.  The
AGN character of both nuclei is revealed by the detection of absorbed,
hard, luminous X-ray emission and of two strong neutral Fe-$\alpha$
lines. Extended X-ray emission components are present, changing their
rich structure in dependence of energy. The close correlation of the
extended emission with the optical H-$\alpha$ emission of NGC 6240, in
combination with the softness of its spectrum, clearly indicates its
relation to star-burst-driven superwind activity.}  }
\label{fig:ngc6240}
\end{figure}

Mergers occur either in gas-poor environments where collisionless/stellar
processes are at play, or/and in gas-rich environments where
gas-dynamical processes assist the black hole inspiral.
Observations seem to reflect this dichotomy: accreting black hole
pairs are seen in elliptical galaxies devoid of cold gas, as well as
in ultra luminous infrared galaxies (ULIRGs), prototypes of gas-rich
mergers between spirals. If accretion is triggered along the course of
the merger, black holes can give origin to a rich phenomenology, that
goes from dual to binary AGN activity, in the radio-loud or
radio-quite mode, according to their dynamics, habitat and merger type.

\bigskip
With {\it dual} AGNs we will indicate, in the following, dual black holes
that are simultaneously active during a galaxy-galaxy encounter, but are still at
large relative separations. In this case they are not physically bound
in a binary and so, their dynamics is determined by the gravitational
potential of the interacting system.  With {\it binary} AGNs we
will indicate active black holes that are at close enough distances
that they form a Keplerian system in a merger remnant. {\it Recoiling}
AGNs have been also hypothesized following the recognition that
asymmetric gravitational wave emission leads to large recoil
velocities: they should appear as off-center AGNs accreting from a
small punctured disk$^{45-47}$.

\bigskip

\subsubsection {Dual AGNs}

In our local universe, 3C 75 is a well studied double radio source at
the center of the galaxy cluster Abell 400$^{48-51}$. Figure 1 shows
the radio and X-ray images of the composite source, and of the host
galaxies in the optical. 
Evidence for interaction between the radio jets suggests that the
sources are members of a system undergoing a large scale merger.  {\it
Chandra} observations have confirmed the presence of two X-ray nuclei
7 kpc apart, and of diffuse thermal emission along the radio streams,
indicating that their proximity is not an artifact of a projection
effect.  3C 75 is not the only case known of a dual AGN: PKS 2149-158
A and PKS 2149-158 B in Abell 2382 is a dumbbell system$^{52}$ hosting two
radio cores with extended twin-jets, observed at a projected distance
of 16 kpc. The jets emanate from two ellipticals of comparable
luminosity and show signs of interaction with the intra cluster medium. Only
through dedicated optical and X-ray follow-up observations it will be
possible to prove that the underlying galaxies are in close physical
interaction.

\bigskip

Dual AGNs have been recently discovered inside the dusty, obscured
environment of at least two ULIRGs whose infrared emission is powered
by an intense star-burst.  The most compelling is the case of NGC
6240$^{53-61}$, an on-going gas-rich merger between two spirals, shown
in Figure 2: {\it Chandra} images have revealed that presence of two
bright nuclear X-ray sources whose spectral energy distribution is
consistent with being two accreting massive black holes (of $0.7$ and
$2.4\times 10^9\, \msun$$^{59}$), seen at a projected distance of 700 pc$^{61}$.
Each nucleus is at the center of a stellar disk in rotation whose
dynamics implies the presence of more than $10^9\,\msun$ of stars$^{57}$.
Kinematic evidence exists that the two nuclei are not physically
bound yet, and that the system is caught in the immediate proximity of
its first close passage. A massive rotating disk of cold gas together
with gas filaments fill the space in between the two active nuclei$^{57}$.
Mrk 463 is the second case$^{62}$: this ULIRG shows clear morphological signs
of an interaction between two gas-rich galaxies, and two luminous
obscured X-ray nuclei are observed at a projected distance of 3.8 kpc.
The interacting star-burst system Arp 299 is a potential third
candidate of a dual AGN$^{63-67}$, due to the presence of a highly obscured
X-ray nucleus (in the galaxy member NGC 3690) and possibly of a second
less luminous one (in IC 694), at a projected distance of 4.6 kpc.
Dynamical mass estimates indicate that the black holes in Arp 299 are
nested in massive ($\gsim 10^9\msun$) nuclear gaseous disks of $\lsim
50$ pc in size.

\bigskip

\begin{figure}
\centering{
\resizebox{6cm}{!}{\includegraphics{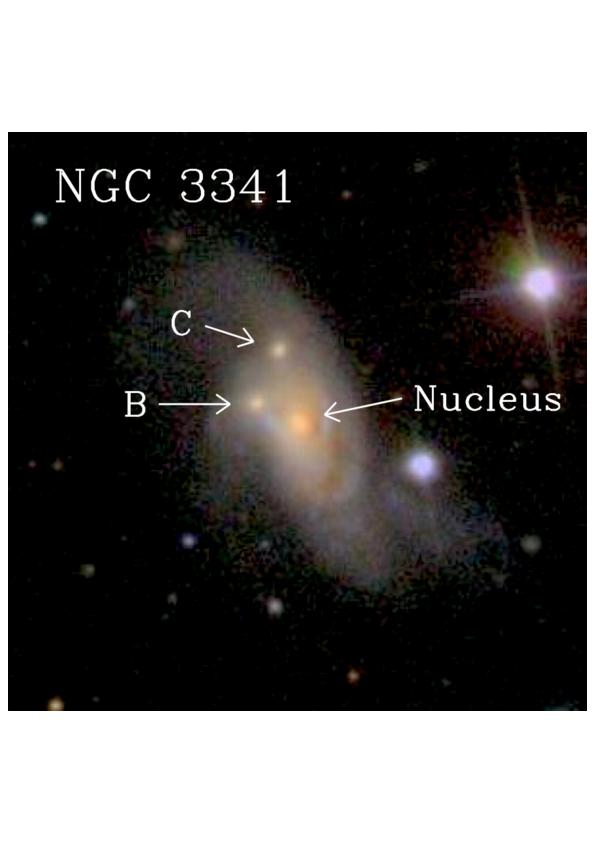}}\\
\caption{SDSS color-composite image of NGC 3341 in the filed of view of 2'x2'$^{68}$.
NGC 3341 is a disturbed disk galaxy undergoing a minor merger with two dwarf companions. Objects B and C
are located at projected separations of 9.5" and 15.6" (5.1 and 8.4 kpc) from the primary nucleus. Object B hosts an active nucleus. } 
}
\label{fig:ngc3341}
\end{figure}

While the above cases refer to dual AGNs in galaxies of comparable
mass (i.e. major mergers), dual AGNs in minor merger are still
difficult to observe.  Despite this, a system of two offset nuclei has
been discovered in the disturbed disk galaxy NGC 3341. The two offset
nuclei have projected distances of 5.1 and 8.4 kpc from the primary
center$^{68}$.  One of the two offset nuclei (object B) shows a Seyfert
2 spectrum, typical of an accreting black hole.  This unusual system,
shown in Figure 3, can be viewed as a minor merger in action with two
dwarf galaxy satellites falling onto the more massive primary.  As the
primary nucleus indicates a possible low-level of AGN activity typical
of a LINER, NGC 3341 may be the first candidate dual AGN system in a
composite, triple minor merger.

\subsubsection {Binary or recoiling AGNs}

The existence of binary AGNs is still debatable at observational
level, and four cases deserve attention.  The first refers to the
elliptical galaxy 0402+379$^{69-72}$.  VLBI (high spatial resolution)
observations have offered the view of two compact flat-spectrum radio
sources C1 and C2 at a projected separation of only 7 pc$^{70}$.  Two
jets depart from C1, while C2 (initially interpreted as a bright knot
along one jet) shows signs of variability that are a distinguishing
feature of an active black hole.  Thus, two massive black holes are
powering the radio-emission, and are likely to form a real {\it
binary} in a passively evolving elliptical. 

\bigskip
The second is the case of OJ 287, a BL Lac object present on
photographic plates since 1891.  Its complex optical light curve shows
a periodic variation of $\sim 12$ yr suggesting that the engine is a
massive binary black hole with the smaller (secondary) orbiting the
larger hole with an observed $12$ yr orbital period$^{73-75}$.  Maximum
brightness is obtained when the secondary black hole, moving on an
inclined orbit, plunges into the accretion disk that surrounds the
primary; a second enhancement in brightness is detected a year later,
interpreted as perturbed gas falling onto the primary.  For OJ 287,
the primary (secondary) black hole has an estimated mass of $2 \times
10^{10}\msun$ ($2\times 10^{8}\,\msun$) implying an orbital decay
timescale via emission of gravitational radiation of only $\gsim 10^{5}$ yr.

\bigskip
 
The third candidate is the quasar SDSS 092712.65+294344$^{76}$.  No
periodicities in the light curve are observed in this source, but a
systematic shift of 2650 $\kms$ in its composite set of emission
lines.  SDSS 092712.65+294344 has been presented in literature as
first case of recoiling black hole$^{77}$: in this interpretation, the
set of very narrow emission lines at the redshift of the galaxy host
result from illumination of central gas by the active black hole.
Moving with line-of-sight speed of 2650 $\kms$ away from the galaxy
core, the black hole that drags its own accretion disk is responsible
of the blue-shifted broad line emission resulting from gas clouds
bound to the black hole itself. This picture is challenged by the
presence, in the optical spectrum, of a third set of narrow lines at
the same redshift of the broad line system which do not find a
clear-cut interpretation. In the recoiling hypothesis, tenuous gas
moving with the black hole is responsible to the emission.  However,
to a closer scrutiny, this quasar offers at least two alternative
interpretations. SDSS 092712.65+294344 may host a binary whereby a
secondary (lighter) black hole, orbiting around a more massive
primary, is fed by matter flowing from a circum-binary disk in close
tidal contact$^{78,79}$. In this picture, the narrow lines that
define the rest frame galaxy would still result from the central
illumination offered by the binary, the broad line blue-shifted set
from gas clouds bound to the orbiting secondary, and the narrow
blue-shifted lines from rarefied gas around the secondary black hole
that has excavated a gap in the surrounding disk (as described in
Section 5.4).  Under this hypothesis, the mass of the primary has been
estimated $\sim 2\times 10^9\,\msun$, the mass ratio between the two
black holes $\sim  0.3$, and an orbital semi-major axis $\sim 0.34$
pc, corresponding to an orbital period of 340 yr and $t_\gw\sim
3\times 10^9$ yr.  The binary hypothesis is preferred over the
recoiling hypothesis being one hundred times more probable than the
ejection hypothesis$^{79}$. Furthermore, it can be directly tested by
searching for systemic Doppler changes in the velocity due to orbital
motion$^{78}$.  A third alternative is that SDSS 092712.65+294344 is
the composite system of two active AGNs co-present inside two galaxies
that are in the verge of colliding at high relative speed in a galaxy
cluster$^{80,81}$.

\bigskip 
The four and last candidate is the quasar SDSS
J153636.22+044127.0$^{82}$ discovered recently in a systematic search
of quasars with anomalous spectra, among 17.500 sources with redshift
$\lsim 0.7$ in the SDSS DR7 release. The quasar shows two clear broad
line emission systems separated in velocity by 3500 $\kms$, plus a
third unresolved absorption line system at an intermediate velocity
(interpreted as emission from intervening material along the line of
sight).  Boronson \& Lauer$^{82}$ interpreted this object as a binary system
of two black holes with their own emission line system. 
\footnote{After submission of this review, 
new observations, in the optical and radio bands, 
are hinting for interpretations alternative to the binary black hole
hypothesis.
Chornock et al. (arXiv:0906.0849) ascribe the peculiar features of this quasar 
to a single massive black hole with properties common to the class of double-peaked emitters. Wrobel and Laor (arXiv:0905.3566) observed
SDSS J1536+0441 using VLA discovering two radio sources separated by 5 kpc within the
quasar optical localization region.  Decarli et al. (ATel n.2061)  and Lauer and Boroson (arXiv:0906.0020) discovered the presence of a companion galaxy 
in correspondence of the second radio source, and most recently, Decarli et al. (submitted to
ApJL) 
found evidence that the quasar lives in a galaxy cluster.
These observations leave open the possibility
that the source is a dual quasar, and only through high angular-resolution
spectroscopy, it will be possible to disentangle the nature of SDSS J1536+0441.}

\subsection{The quest for massive black hole binaries in ellipticals}

Elliptical galaxies are believed to 
be the end product of multiple mergers$^{83-85}$, though the exact sequence of events
that shape ellipticals as they appear today is still uncertain.
Ellipticals show remarkable regularities in their smooth stellar light profile: their outer surface brightness $I(r)$ 
follows a S\'ersic  profile $\log I (r)\propto r^{1/n}$ over a wide mass-dynamical range. The fit is so remarkable that departures from this regularity  
has been recently used as diagnostic tool for tracing back their formation history$^{86}$. 

\bigskip
Ellipticals appear to come in two distinct varieties$^{86,87}$: bright (giant) ellipticals 
have S\'ersic index $n>4$ and show cores, i.e. {\it missing-light} inside a radius 
marking the dividing line between a flatter inner surface brightness profile and  
the S\'ersic profile at outer radii; the less bright, less massive (normal) ellipticals
have instead $n\lsim 4$, and show {\it extra-light} at their center
relative  
to the inner extrapolation of the outer S\'ersic profile.
Missing or extra light are features that underscore a dichotomy 
that involves not only the central kpc of the galaxy but many global 
structural and physical properties$^{88}$. 
Core (i.e. missing-light) giant ellipticals are essentially non-rotating, anisotropic and
modestly triaxial; stars are very old and if gas is present, it  is {\it hot}
and X-ray emitting. Often giant ellipticals host strong radio sources.
By contrast, core-less (i.e. extra-light) ellipticals  
are rotationally supported and have disky-distorted isophotes.
Rarely contain hot X-ray emitting gas and a  radio source. 
Both local and  global features thus appear to record memory
of a different merger history. A number of authors$^{86,89,90}$
has suggested recently 
that missing-light (giant) ellipticals are shaped by (re-)mergers of {\it  gas-poor}
galaxies, while  extra-light ellipticals are the result of a  {\it gas-rich} merger
producing  an extra stellar component in the central regions of the galaxy $^{89}$.
Extra-light in normal ellipticals is suggestive that {\it cold} gas has played a critical role 
in the build up of the galaxy: 
gas can radiate and fall in the central regions of the merger remnant converting 
a sizable fraction of its mass into stars, in violation of  Liouville's theorem. 
Missing light$^{90}$ instead does not find an easy explanation:  
collisionless mergers preserve the density profile of the progenitor galaxies 
and do not enhance significantly the velocity dispersion, according to Liouville's theorem. Thus, the presence of missing-light requires extra physics.

\bigskip

Core (missing-light) galaxies and core-less (extra-light) ellipticals
generally host a massive black hole in their nucleus, and an
interesting hypothesis has been advanced. That in a gas-poor
environment, missing-light results from a dynamical process involving
a {\it binary} black hole. A first possibility is that a
stellar core forms in response to repeated three-body interactions
between the binary and a star passing by$^{91}$: impinging on the
binary from low angular momentum orbits stars are scattered back with very
high velocities and leave the nucleus.  The binary that decays excavates with time the
central region of the galaxy producing a stellar core, emptied of
light.  The missing light correlates with the black hole mass 
closely$^{87,92,93}$and with the number of successive mergers
necessary to explain the stellar mass deficit$^{93,94}$.
A second possibility that may operate in succession is core
formation due to the recoiling black hole born after binary
coalescence.  Gravitational radiation not only carries away energy and
angular momentum: depending on the black hole spin direction relative
to the orbital plane, radiation transports away linear momentum, as
waves are emitted anisotropically.  The relic black hole, as already
mentioned in Section 1.2, receives a recoil velocity that can be as
large as 4000 $\kms$.  The instantaneous removal of the black hole via 
radiation recoil generally results in the rapid formation of a stellar
core that expands due to the sudden change in the gravitational
potential, now lacking of a massive central body, and to the
subsequent deposition of kinetic energy as the black hole returns
toward the center of the galaxy$^{95-97}$ 
(the effect being the largest when the
hole is ejected on a return orbit with a velocity just below escape).
These ideas radical and ad hoc have become nowadays two mainstream
possibilities.
   
\bigskip
But, how can core excavation be prevented in core-less ellipticals? 
A possibility is that gas has played a critical role not only in creating new 
additional light but in  
accelerating binary decay (if a binary is present) under the action of
gas-dynamical torques.  The working hypothesis is that fast orbital decay controlled by gas prevents core formation; spin-orbit alignment which is favored
in a gas rich environment$^{98}$, brings the two holes 
in a configuration, prior to coalescence, of minimum recoil,
thus preventing expansion of the stellar core at the time of formation of the new, relic
hole.

\bigskip 

>From the above considerations, black hole binaries seem to be linked in 
fundamental ways to the dynamics of the stellar and gas components of 
galaxies, and this motivates strongly  the interest on binary black holes not
only as sources of gravitational waves, but as key systems  
in the context of galaxy formation.  

\section {Black hole dynamics in galaxy mergers}

Mergers occur mainly in two varieties: they may involve nearly equal
mass galaxies, and these are referred to as {\it major} mergers, or
unequal mass galaxies, i.e. {\it minor} mergers.  Both may involve
either gas-rich or gas-poor galaxies so that black hole dynamics has
no unique outcome but differs in relation to the merger type and
environment.  Exploring these diversities in a self-consistent
cosmological scenario is currently a challenge, and thanks to recent
advances in numerical simulations, it is now possible to address a
number of compelling issues.  Galaxy mergers cover cosmological
volumes ($\lsim 100$ kpc aside), whereas black hole coalescences probe
exceedingly small volumes from $\lsim 1$ pc down to $a_\gw\sim
10^{-3}$ pc and to the horizon scale ($10^{-7}$ pc, for $\sim 10^6
\, \msun$).  Thus, tracing the black hole inspiral in a merger requires
simulations with gravitational/hydrodynamical force resolution
covering more than ten orders of magnitude in length, and a variety of
different physical processes, often not scale free.

\bigskip
Following a merger, how can black holes reach the gravitational wave
inspiral regime?  The overall scenario has first been outlined in a
seminal study by Begelman, Blandford \& Rees $^{99}$, 
in the context of stellar systems.  Four main phases characterize the 
dynamical evolution
of the black holes: (I) the pairing
phase P in which dynamical friction against the stellar/dark matter
background is acting on each black hole causing their sinking; (II) phase B when the holes
dynamically couple to form a {\it binary} and continue
to sink due to the large scale drag of dynamical friction; (III) phase H
in which the binary hardens via 3-body scatterings off single stars;
(IV) phase GW, the last, when gravitational wave back-reaction becomes
important to drive the inspiral down to coalescence.

\bigskip

Early studies explored the phase of pairing P, simulating the collisionless merger
of spherical halos $^{100,101}$.  
Governato, Colpi \& Maraschi$^{100}$ in particular first noticed that
when two equal mass halos merge, the twin black holes nested inside
the more massive nuclei are dragged rapidly toward the center of the remnant
and form a close pair.  The drag,
initially, is not acting on the individual holes but on the massive
stellar-cusp in which they are hosted. Only when the merger is in its
advanced stage, the black holes behave as independent point masses and
continue to lose orbital angular momentum under the action of dynamical
friction.  They further noticed that the situation reverses in unequal
mass (minor) mergers, where the less massive halo, tidally disrupted
along the course of the encounter, leaves its {\it naked} black hole
wandering in the outskirts of the main halo.  Thus, depending on the
halo mass ratio and internal structure of the interacting galaxies,
the transition from phase P $\to$ B can be at some point
prematurely aborted or severely delayed.  Similarly, the transit from phase B $\to$ H $\to$ GW
is not always secured even in major mergers (where also the black
holes are expected to have comparable masses), as the stellar content
in the immediate vicinity of the black holes may not suffice to harden
the binary down to a separation where gravitational wave driven
inspiral sets in$^{101-104}$. 
All these issues will be discussed critically below. 

\bigskip
N-Body and Monte Carlo merger trees simulations$^{105-113}$ describing
the assembly of galaxies show that unequal mass mergers are common
along cosmic history. As black hole masses, in the local universe,
correlate tightly with the mass of the host$^{1-4}$ ($M_\bh\sim 
10^{-3}M_{\rm halo}$), this implies a black hole mass spectrum
ranging from $10^5\,\msun$ up to $10^9\,\msun$ and mass ratios
$q_\bh\sim 0.01-1.$ In literature, these intervals have not been
explored systematically and main emphasis has been given to equal-mass
mergers, given the heavy computational demand, and the impossibility,
in gas-rich mergers, of using self-similar solutions, owing
to the presence of dissipative components.  While black hole masses in
the interval $10^5-10^7\,\msun$ are of interest for {\it LISA}
astrophysics, black holes with $M_\bh>10^8\,\msun$ are of key
importance for the Pulsar Timing Array Experiment (PTA), due to their
different frequency of operation. While the "light" {\it LISA} black
holes carry information on how/where they formed along the course of
galaxy assembly $^{110}$(up to $z\sim 20$), the "heavy" PTA black holes$^{34}$ trace
mergers among very massive galaxies present in the already evolved
universe, at redshift $z\lsim 1$.

\subsection{Capturing  basic physics}

\subsubsection {Dynamical friction}

Dynamical friction is the main cause of braking of the black hole orbits during 
the process of pairing, and acts individually on each hole. 
This drag arises in response to the gravitational perturbation excited by the black hole
on the background of dark matter and/or stars. In the original formulation of
Chandrasekhar$^{114}$, the drag is the collective result of momentum exchange
between the massive perturber $M_\bh$ 
(the black hole in our case) and every single particle of
the background. In a collisionless system of uniform 
mass density $\rho_*$ and isotropic velocity distribution,  
the force on $M_\bh$, acting against its velocity vector $\bf V$  is   
\begin{equation}
{\bf F}_{\rm DF}
= -4\pi\ln\Lambda G^2 M^2_\bh \rho_* 
\left [{\rm{erf}} \left({V\over \sqrt {2}\sigma}\right)-\left (\sqrt{{2\over \pi}}{V\over \sigma}\right )
\exp\left(-{V^2\over2\sigma^2}\right)\right ] {{\bf V}\over V^3}.
\end{equation}
The gravitational drag that involves only those particles moving more
slowly than $V$ (the term in squared brackets), is maximum when the
velocity $V$ of the perturber nears the 1D velocity dispersion
$\sigma$, as the force declines linearly with $V$, and as $V^{-2}$, in
the low ($V\ll\sigma$) and high ($V\gg \sigma$) speed limit,
respectively.  The Coulomb logarithm $\ln \Lambda\sim \ln (b_{\rm
max}/b_{\rm min})$ in equation (5) expresses the non-locality of the
drag as it comprises the whole range of impact parameters relevant for
the exchange of momentum in the interaction of the stars with $M_\bh.$
Since gravity is a long-range force, $b_{\rm max}$ is close to the
size $L$ of the collissionless background, while $b_{\rm
min}=V^2/GM_\bh$ is the impact parameter for a large-scattering angle
gravitational interaction.  Dynamical friction expresses the tendency
of gravitating systems to reach energy equipartition, so that any
excess of kinetic energy (that carried by the hole) is distributed
among all particles.  Restricted to a static, infinite, isotropic and
homogeneous collisionless medium, Chandrasekhar's formula assumes that
the stars/dark-matter particles move along unperturbed trajectories
that are straight lines, thus neglecting the self-gravity present in
real backgrounds.  Despite this limitation, the formula has been often
used to estimate sinking times also in inhomogeneous self-gravitating
backgrounds evaluating $\rho_*$ and $\sigma$ at the current, i.e.
local position of the perturber, treating $\ln\Lambda$ as adjustable
parameter.

\bigskip
Many approaches has been developed to overcome the limits of the local
approximation, including the theory of linear response TLR$^{115,116}$.
In TLR, dynamical friction is viewed as a direct manifestation of the
fluctuation-dissipation theorem$^{117,118}$.  In this interpretation, the
fluctuations of the two-body force between the massive perturber
$M_\bh$ and any particle $m_*$ add collectively to give a
non-vanishing drag. In a medium characterized by a distribution
function $f({\bf r},{\bf p})$, the components of the force can be
expressed as the sum over time and phase-space ({\bf r},{\bf p})
\begin{equation}
F^a_{\rm TLR}=G^2M^2_\bh\int_{-\infty}^t ds\int d_3{\bf r}d_3{\bf p}
\,\,{{\partial f\over \partial p^b}}\,\,
T^{ba}
\end{equation}
of a suitable self-correlation tensor 
\begin{equation}
T^{ba}=Nm_*^2\,\,{ r^b_\bh(s)-r^b\over \vert {\bf r}_\bh(s)-{\bf r}\vert^3}\,\,{
r^a_\bh(t)-r^a(t-s)\over \vert {\bf r}_\bh(t)-{\bf r}(t-s)\vert^3}
\end{equation}
correlating the two-body gravitational force between the black hole (in ${\bf r}_\bh$)
and any individual star (in $\bf r$) at time $t$ with that at
the earlier time $(t-s)$.  This formula captures the complexity of the
physics leading to the orbital decay of a mass $M_\bh$ in a generic
self-gravitating background.  The drag force at time $t$ keeps {\it
memory} of the previous history of the composite system, as its
strength depends on the dynamics of the perturber and of the particles
at earlier times as determined by the unperturbed Hamiltonian.  Applied
to selected spherical environments$^{116}$, TLR has been powerful to prove
that the stellar response embodied in $T^{ba}$ remains correlated with
the perturbation for a time shorter than the typical orbital time of
stars and that the force arises preferentially from those stars closer
to the perturber, due to the inhomogeneity of the background (distant
lower density regions contribute less to the drag).  Tested against
high resolution numerical simulations$^{116}$, TLR has been useful also in
showing that in the interplay between loss of energy and angular
momentum, eccentric orbits do not circularize under the action of the
drag force, but remain eccentric along the course of the decay
implying shorter sinking times.  In a singular isothermal sphere with
1D velocity dispersion $\sigma$ and density profile
$\rho(r)=\sigma^2/[2\pi G r^2],$ TLR predicts a sinking time,
expressed in terms of the circularity $\epsilon$ (defined as the ratio
between the angular momentum of the current orbit relative to that of
a circular orbit of equal energy)
\begin{equation}
\tau_\df=1.2 {r_{\rm cir}^2V_{\rm cir} 
\over \ln(M_{\rm halo}/M_\bh) \,GM_\bh}\epsilon^{0.4},
\end{equation}
where $r_{\rm cir}$ and $V_{\rm cir}$ are the initial radius and
velocity of the circular orbit with the same energy of the actual
orbit, and $M_{\rm halo}$ is the mass of the dark matter and stars
within $r_{\rm cir}.$ Equation (8) indicates that a massive black hole
can sink at the center of the sphere within a time $\sim 10^{8}$ yr,
if released during the merger at a distance of $\sim 100$ pc:
\begin{equation}
\tau_\df\sim  
5\times 10^8 \left ({5\over \ln[M_{\rm halo}/M_\bh]}\right ) \left ({r_{\rm cir}\over 300 {\rm pc}}\right )^2
\left({V_{\rm cir}\over \sqrt{2} \times 100 \,\kms}\right)
\left({10^6\,\msun\over M_\bh}\right)\epsilon^{0.4}\,\rm yr. 
\end{equation}

\bigskip

Dynamical friction acts also in collisional fluids and arises again
from the gravitational pull between the perturber and its density wake
excited in the ambient medium.  In the steady-state limit and for {\it
supersonic} motion, the drag on a massive perturber moving with
velocity $\bf V$ across a homogeneous fluid with density $\rho_{\rm
gas}$ and sound speed $c_{\rm s}$ reads
\begin{equation}
{\bf F}^{\rm gas}_{\rm DF}=-4\pi \ln\left [ 
{b_{\rm max} \over b_{\rm min} }
{ ({\cal M}^2-1)^{1/2}\over {\cal M}}\right ]
G^2\,M_\bh^2\rho_{\rm gas}
{{\bf V}\over V^3}, \,\,\,\,\,\,\,{\rm for}\,\,\,\,\,\,\,{\cal M}>1
\end {equation}
where ${\cal M}=V/c_{\rm s}$ is the Mach number.  The deceleration
results, in this regime, by the enhanced density wake that lags behind
the perturber and that is confined in the narrow cone, the Mach cone, whose
surfaces of constant density are hyperboloids$^{119}$.  The main result
is that the gaseous drag is enhanced for supersonic motion (with
${\cal M}\sim 1-2.5$) compared to Chandrasekhar's formula (for
$\rho_*=\rho_{\rm gas}$) by a factor $\sim  2$ hinting for shorter
timescales when dynamical friction occurs in gaseous
backgrounds$^{119}$.  A black hole can also accrete while sinking and a
drag, comparable in strength to the large-scale force ${\bf F}^{\rm
gas}_{\rm DF}$ arises due to accretion: ${\bf F}^{\rm gas}_{\rm
BHL}=-{\dot M}_{\rm B}{\bf V}$ where ${\dot M}_{\rm B}$ is the
Bondi-Hoyle-Littleton accretion rate so that

\begin{equation}
{\bf F}^{\rm gas}_{\rm BHL}=-4\pi \lambda G^2 M^2_\bh{\rho_{\rm gas}\over(V^2+c_{\rm s}^2)^{3/2}}{\bf V}
\end{equation} 
(with $\lambda\lsim 1$), valid for any $\cal M.$ 

\bigskip
  
The {\it subsonic} case is instead more elusive, because sound waves
can propagate both down-wind and up-wind so that the drag, in this low
speed limit, can be much weaker than in the collisionless case.  The
drag is exactly zero in a homogeneous infinite medium due to the
front-back symmetry of the perturbed density distribution present in
any stationary solution.  However this result does not rigorously hold
if one performs a finite time analysis$^{119}$.  If the perturber
triggers the disturbance at time $t=0$ and moves subsonically along a
straight line, the symmetry is broken as long as $(c_{\rm s}+V)t$ is
smaller than the size of the medium, resulting in a finite drag $ {\bf
F}^{\rm gas}_{\rm DF}=-(4/3)\pi G^2\,M^2_\bh \rho_{\rm gas} {{\cal
M}^3}{\bf V}/ V^3\propto M_\bh^2\rho_{\rm gas}\,{\bf V}/c_{\rm s}^3$
for ${\cal M}\ll 1.$

\bigskip

Equations (5-11) set the guidelines for our understanding of dynamical
friction.  As the drag $F_\df$ is proportional either to $\rho_*$ or
$\rho_{\rm gas},$ black holes find their preferred way toward
coalescence in backgrounds as those created in dissipative
mergers.  Gas can play an important role there: subject to 
instabilities and shocks, it  can cool and 
control preferentially the inspiral creating conditions were  $\rho_{\rm gas}>\rho_*$. 
$F_\df$ is also a sensitive function of $V/\sigma$
or ${\cal M}=V/c_{\rm s}$ and is large when $V/\sigma$ and/or ${\cal
M}\sim 1$.  Nevertheless, this basic formulation can not capture
important details of the process. In a real merger:

\medskip
$\bullet$ The underlying gravitational field is fluctuating violently; 

$\bullet$ Prior to the encounter, the black holes
are often inside 
massive power-law stellar cusps 
and thus have effective mass $M_{\rm BH,eff}$ larger than their own mass:

$\bullet$ Tides strip stars in the cusp and  $M_{\rm BH,eff}$ changes 
long the course of the merger, as $M_{\rm BH,eff}\to M_\bh$; 

$\bullet$ Stars and gas are co-present, but often 
have different geometries and so, they respond 
differently to the perturbation induced by the black hole: in rotating backgrounds 
the gravitational pull of $M_\bh$ creates spiral patterns  
that can not simply be described by a back-stream density wake; 

$\bullet$ The density wakes excited by the two black holes
may start to self-interact and overlap as the hole's separation decreases.

$\bullet$ When the process of pairing under dynamical friction is
sufficiently advanced, the black holes form a binary, and at this
stage the interaction changes character.  In presence of a
collisionless (stellar) background, it is the interaction of
individual stars that drives binary evolution, while in a collisional
fluid, gravitational and viscous torques play a key role.

$\bullet$ The black holes may start to inject energy and momentum in the medium 
modifying the underlying density, and in turn the back-reaction force on their orbit. 

\medskip

\subsubsection{From Pairing to Hardening: key scales}

\bigskip

If, in the simplest approximation, the collisionless background is 
described as
a singular isothermal sphere, two black holes form a binary when their separation 
$a$ falls below the   
distance 
$a^*_{\rm binary}$ at which the mass in stars/dark matter enclosed  
within their relative orbit is equal to the total 
mass $M_\bht$ of the binary: 
\begin{equation}
a^*_{\rm binary}\sim {GM_\bht \over 2\sigma^2}\sim 0.2 \left ({M_\bht\over 10^6\,\msun}\right )\left ({100 \, \kms\over \sigma}\right )^2\,\,{\rm pc}.
\end{equation}
At $a\sim a^*_{\rm binary}$, the hole's relative circular velocity 
equals the circular velocity of the stars, $\sqrt{2}\sigma.$
Dynamical friction continues to drive the inspiral, but as soon as $V_{\rm cir}$
increases above $\sqrt{2} \sigma,$ 
the interaction between the binary and the stars from
collective becomes individual, and 
the drag no longer shows the clear dependence on $M_\bht$.
{\it Scattering off the binary}
by {\it single stars} becomes the main
process for extracting binding energy and angular momentum from the orbit.
When a star is impinging on the binary, it is shot out at an average velocity comparable 
to the binary's orbital velocity $V_{\rm cir}$ (eq. 4).
If $V_{\rm cir}$ exceeds the 3D velocity dispersion $\sqrt{3}\sigma$, stars
are ejected away from the galaxy nuclues
with very high speeds, and this 
occurs as soon as  the binary reaches the hardening radius$^{120}$,
defined as the
binary separation at which the binding energy per unit mass exceeds $(3/2)\sigma^2$: 
\begin{equation}
a^*_\hard=
{G\mu_\bh\over 3\sigma^2}\sim  0.13{q_{\bh} \over (1+q_{\bh})^2}\left ( {M_\bht\over 10^6\,\msun}\right )
\left (\,{100 \, \break \kms\over \sigma}\right )^2\,\,\rm{ pc}.  
\end {equation}
Below $a^*_{\hard},$ stars 
leave the galaxy's core carrying away
a tiny fraction of the binary energy, approximately 
$\sim  (3/2)Gm_*\mu_\bh / a,$
 where $\mu_\bh=M_\bht q_\bh/(1+q_\bh)^2$ is the reduced mass of the binary$^{120}$. This corresponds to an increase of the binary's binding energy $E_{\rm B}$ and thus a decrease in
the semi-major axis per scattering of the order of  
$\delta a/a=\delta E_{\rm B}/E_{\rm B}\sim m_*/M_\bht.$ 

According to this loss, the binary continues to harden transiting 
from B $\to$ H, and from H $\to$ GW, and 
to accomplish this, 
it needs to eject  a mass in stars 
\begin{equation}
M_{\rm ej}\sim  {1\over 3}\mu_\bh \ln\left ({a^*_{\rm hard}\over a_\gw}\right )\sim 
(10-20) \mu_\bh
\end{equation}
for a typical galaxy's nucleus$^{120,121}$.
Whether the binary will eject this mass, and complete the inspiral down
to the gravitational wave domain will be discussed in Section 3.

\bigskip

If the black holes move inside the gaseous background of a nuclear
disk that forms in a merger, we can introduce the scale below which
they bind $a_{\rm binary}^{\rm gas}:$ to this purpose, we consider as
reference model a Mestel disk, i.e. a rotationally supported,
self-gravitating disk characterized by a surface mass density profile
$\Sigma=\Sigma_0(R_0/R)$ where $R$ is the projected radial distance
and $\Sigma_0$ the density at the scale radius $R_0.$ The disk, in
differential rotation $\Omega_{\rm disk}(R)=V_{0,\rm rot}/R$ with
constant $V_{0,\rm rot}$, has a total mass $M_{\rm disk}=2\pi\Sigma_0
R_0^2$. The black holes form a binary whenever the disk mass enclosed
within $a$ equals $M_\bht$:
 \begin{equation}
 a^{\rm gas}_{\rm binary}=  R_0 {M_{\bht}\over M_{\rm disk}}\sim  1 
 \left ({R_0\over 100 \,{\rm pc}}\right )
 \left ({M_\bht\over 10^6\,\msun}\right )
 \left ({10^8\,\msun\over M_{\rm disk}}\right )\,\,\rm {pc}.
\end{equation}
In a fluid, the binary does not interact with an isolated parcel of gas
since any perturbation is transmitted to the fluid either via shocks or
through acoustic waves. Thus, below $a^{\rm gas}_{\rm binary}$ the
interaction remains collective.  In the transit from B $\to$ H $\to$
GW, the binary continues to exchange angular momentum with the gas and
to deposit energy that can be radiated away, in a manner that will be
described in Section 5.
\bigskip

A key radius that can be introduced to describe the coupling 
of the black hole with its environment, 
is the gravitational influence or Bondi-Hoyle-Littleton radius
\begin {equation}
R_{\rm BHL}\sim {GM_\bh\over (c_{\rm s}^2+V^2_{\rm rel})}\sim 0.4 
\left({M_\bh\over 10^6\,\msun}\right ) \left (
 {(100 \,\kms)^2 \over c_{\rm s}^2+V^2_{\rm rel}}
\right )\,\, {\rm pc}
\end{equation}
(where $V_{\rm rel}$ is the velocity of the black hole relative to the
medium) that indicates the size of the region of dynamical influence
of each black hole on fluid particles.  Thus, as soon as the binary separation drops below
$R_{\rm BHL}$, the gravitational field of both holes is expected
to affect the geometry, dynamics and thermal state of the gas, and key
physical processes for black hole inspiral will be described in
Section 5.

\subsection{Capturing basic physics from key numerical experiments}


\begin{figure}
\centering{\resizebox{10. cm}{!}{\includegraphics{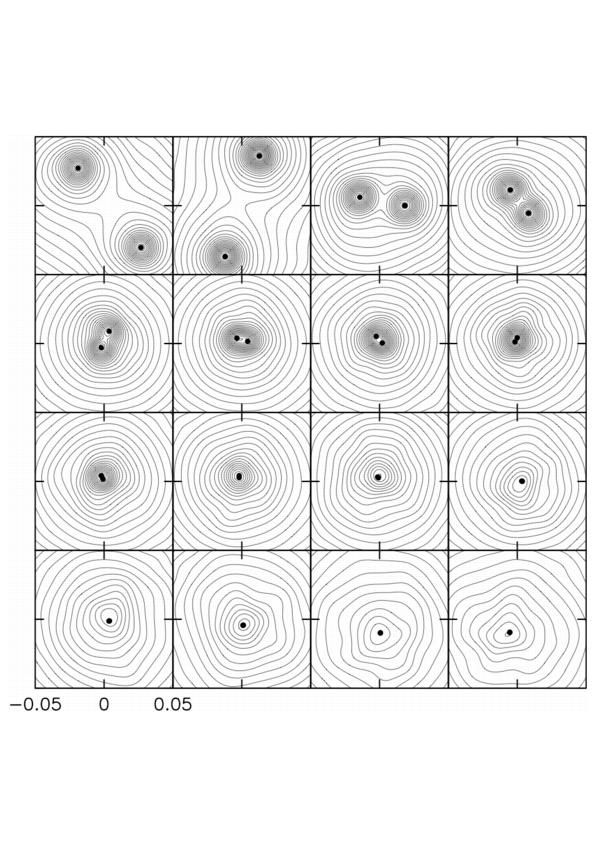}}\\
\caption{Projected density contours 
describing the merger of twin
spherical stellar systems set initially on an elliptical orbit$^{101}$. 
The two black holes (filled circles) 
of total mass $M_\bht=0.01M_{\rm galaxy}$ are at the center of the two stellar cusps 
that spiral-in
under the influence of dynamical friction (first row). The two cusps merge
(in second row) and the black holes form a binary inside a newly regenerated stellar 
cusp whose profile is close to the initial profile.
Three-body scatterings of single stars cause the progressive 
hardening of the binary and the excavation of a stellar
core as stars are ejected above escape velocity. The density of the newly-formed
cusp drops rapidly as the black hole binary transfers energy to the stars. 
}
}
\label{fig:merritt}
\end{figure}

\bigskip
Figure 4 shows the result of a study by Milosavljevic \& 
Merritt$^{101}$ who simulated the idealized merger of two identical
spherical galaxies with twin black holes, combining the tree code
GADGET$^{122}$, used as integrator to trace the early stages of the merger
when the black holes are subject to dynamical friction in the pure
stellar background, to the code NBODY6++$^{123}$, a direct-summation N-Body
code that follows the dynamics of the black holes scattering off
individual stars, in the advanced stages of the merger.
%
%
\begin{figure}
\centering{
\resizebox{8cm}{!}{\includegraphics{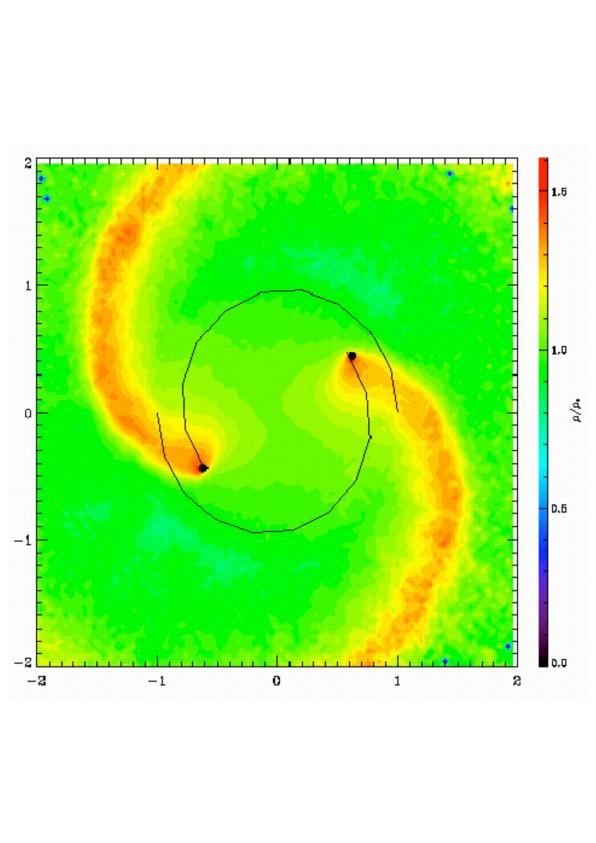}}
\resizebox{8cm}{!}{\includegraphics{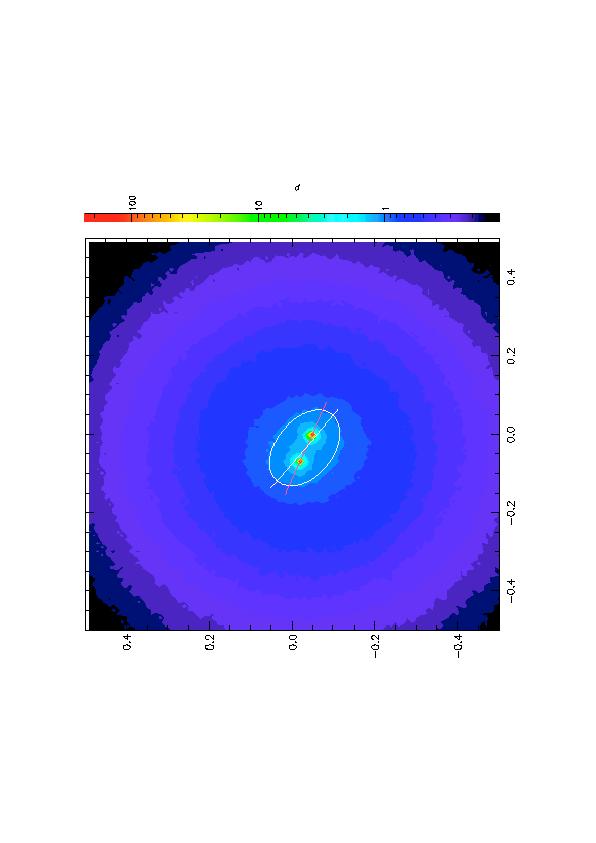}}\\
\caption{Left panel shows the density perturbation caused by twin
black holes orbiting in an isothermal gas clouds (face on view above
the orbital plane).  Black curves represent the individual orbits of
the black holes that spiral in under the action of gas-dynamical
friction.  In this simulation$^{124}$ the mass of the binary is 0.02 of
the mass in the gas-sphere.  Right panel shows the color-coded
density map, in the orbital plane, when the black hole separation
$a<a^{\rm gas}_{\rm binary}.$ The two density peaks in red coincide
with the location of the black holes. An ellipsoidal deformation with
axis misaligned with the respect of the binary axis is present and is
exerting a torque on the binary.  }}
\label{fig:escala}
\end{figure}
As illustrated in Figure 4 (upper panel, first row) the black holes
initially remain closely associated with their stellar cusp. In this
phase black hole pairing is controlled by dynamical friction which is
acting not only on the individual black holes but on the black
hole+star's cusp system. As time proceeds, the two cusps merge into a
regenerated cusp with density comparable to the initial, and the black
holes form a binary (second row).  In the last two rows, the
interaction of the binary with the stars is no longer
collective and is followed with the direct-summation N-Body code. 
Dynamical friction plays no role, and three-body
scattering becomes the main process of binary hardening that continues
at a different peace.  Meantime, stars are ejected and core scouring
starts.  This is illustrated again in Figure~4 (last two rows) where
the density of the newly-formed cusp continues to drop as the black
hole binary transfers energy to the stars. 
The code NBODY6++ can not follow however the {\it long-term} evolution
of the binary as a variety of spurious collisional effects, associated
to the small number of particles $N$ used, leads to an 
unphysical, rapid decay of the binary.  Because of the numerical
discreteness effects associated with approximating a galaxy nucleus
with $N\gsim 10^9-10^{10}$ stars by a model
consisting at best of $N\gsim 10^6$ particles, direct N-Body
simulations appear to be limited to the early stages of phase H.  For
this reason the subsequent evolution will be described in Section 3.

\bigskip
As far as gas-drag is concerned, Figure 5 shows a case-study by Escala et al.$^{124}$
who simulated using GADGET in its SPH implementation, the sinking of
two black holes in a spherical isothermal gas cloud.  The black holes,
initially unbound, excite two prominent density wakes along their
mildly transonic orbit and sink toward the center where they form a
binary. When the large-scale drag by gas-dynamical friction is no
longer effective since $a$ has decayed below $a^{\rm gas}_{\rm
binary},$ the system develops an ellipsoidal deformation resulting
from the superposition of the heads of the wakes.  The asymmetry
present in the wakes combines to make the ellipsoidal gaseous envelope
misaligned with the binary axis, and this is the cause of further
decay.

\section{Black hole dynamics in stellar backgrounds}

There are, at present, no simulations of {\it gas-free} mergers
between galaxies with composite structure (i.e. with a dark matter
halo, a stellar disk, a bulge and a central black hole) that follow in
realism the dynamics of the black holes during phase P $\to $ B.  Most
of the studies start already with a close black hole pair moving
inside the gravitational potential of a stellar equilibrium model, and
follow in great detail the phase of binary hardening by three-body
interactions (phase H), using direct summation codes.  These studies
have revealed that the rate at which the binary hardens depends
sensitively on the properties of the underlying gravitational
potential and on the degree of two-body relaxation present in the
galaxy's model$^{103,120,125}$.

\bigskip

As the fractional energy exchange per scattering, in three-body encounters, 
is $\sim m_*/M_\bht,$
the rate at which the binary decays is $-{\dot {a}}/a\sim
(m_*/M_\bht)n_*\sigma A_\bh\sim\pi G\rho_*a/\sigma$, assuming a
constant flux of stars ($\sim n_*\sigma$) and a binary cross section
$A_\bh\sim \pi a (GM_\bht/\sigma^2)$ corrected for gravitational
focusing.  This defines a hardening timescale $^{120}$
\begin{equation}
 t^*_\hard \sim {\sigma\over \pi G \rho_* a}
\sim 7\times 10^8 \left ({\sigma\over 100 \,\kms}\right) 
\left ({10^4\,\msun\,{\rm pc}^{-3}\over \rho_*}\right )
\left ({10^{-3}\,{\rm pc} \over a}\right ) \,\,{\rm yr}
\end{equation}
which expresses the time it takes the binary to reach the separation
$a$, from large initial distances, inside a fixed stellar homogeneous
background.  In galactic nuclei, $t^*_\hard$ 
determines the initial rate at which the binary hardens. Only
stars on near radial orbits with specific angular momentum $J$ less
than $J_{\rm lc}\sim (GM_\bht a)^{1/2}$ experience the field of the
individual binary components: this condition defines a selected domain
in phase-space, called {\it loss cone}.  As the stellar mass contained
in the loss cone of a typical galaxy is less than $M_{\rm ej}$, the
black hole orbit decays at a much lower rate once the loss cone is
emptied$^{125}$.  Whether the black holes can continue to exchange energy and
momentum with stars then depends on the efficiency with which stars
are re-supplied to the loss cone.  The depletion of the orbital
population inside the loss cone may thus lead to a {\it stalling} in
the decay rate.

\bigskip

\begin{figure}
\centering{\resizebox{10 cm}{!}{\includegraphics{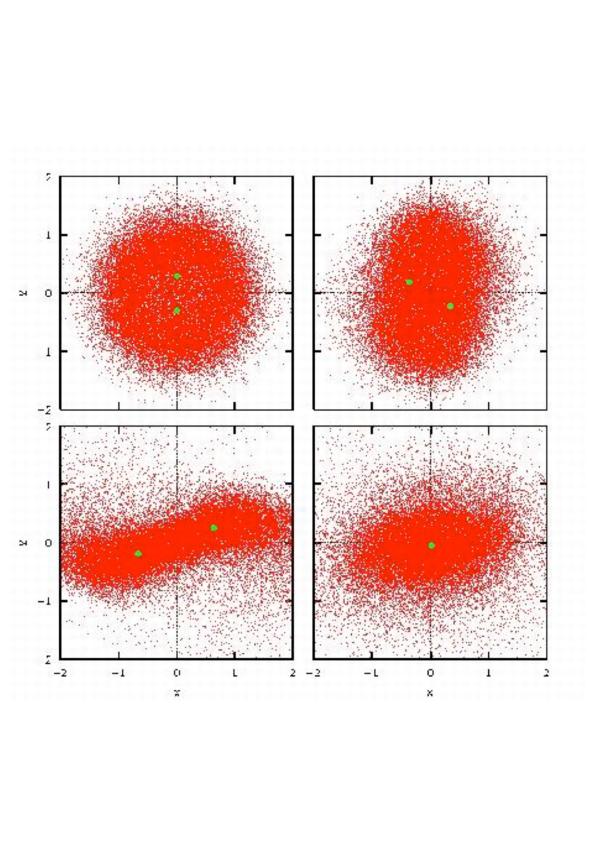}}
\caption{ Evolution of a massive black hole binary (of total mass
$M_{\bht}=0.04 M_{\rm galaxy}$) in a triaxial, rotating galaxy
model$^{128}$.  Snapshots show the positions of the 200K particles of
the simulation, at four selected times.  The twin black holes,
indicated in green, initially come together by falling along the bar
before forming a bound pair. On a longer timescale the binary hardens
at a rate which is found to be independent of the number of particles
used, and sufficiently rapid to allow for coalescence in 10 Gyr, even
in the absence of collisional loss-cone refilling, providing a
potential solution to the final-parsec problem.  } }
\label{fig:berczik}
\end{figure}

Stars are expected to diffuse across the loss cone boundary through
stellar two-body relaxation processes. A small angular momentum
perturbation, excited by a star passing by or by the fluctuating
component of the gravitational potential, can deflect a star with
$J>J_{\rm lc}$ into the loss cone $J<J_{\rm lc}$, and this occurs on
the two-body relaxation time $t_{\rm rel }=t_{\rm cross}N/(8\ln N)$
(where $t_{\rm cross}$ is the crossing through the galaxy of a
typical star). Since $t_{\rm rel}$ is often longer than the lifetime of
the galaxy's nucleus, stellar dynamical processes are unable to
conduct the binary to coalescence, and this theoretical difficulty has
been called the {\it final parsec problem}$^{120,121}$.

\bigskip

This problem has been investigated in depth$^{101-104, 125-132}$, and
the picture that is emerging goes in the direction of reducing the
bridge separating phase H to phase GW, because of the presence of
additional mechanisms$^{125}$.  In the collisionless background of a massive
elliptical where two-body relaxation is unimportant, binary evolution
can continue if stars ejected by the black holes remain, after the
encounter, inside the loss cone: moving on returning orbits, these
stars interact again with the binary for a few times before they are
lost$^{125}$, alleviating the demand on $M_{\rm ej}$.  If re-ejection
happens to be only marginally important, most relevant is the geometry
of the galactic potential.  In triaxial nuclei 
where angular momentum is not a conserved quantity, there
exists a potentially larger population of orbits that can encounter
the black hole binary$^{127}$.  The non-axisymmetric/rotating case in particular
allows for the growth of stellar bars$^{128}$. Torques from bar-like
potentials then create a population of centro-philic orbits that pass
near the center of the galaxy once per crossing time, and that
promptly interact with the binary.  Figure 6 shows, via a sequence of
snapshots, how black holes sink inward along the bar, inside an
unstable strongly rotating stellar spheroids$^{128}$.  A further
mechanism for refilling the loss cone is dynamical relaxation induced
by a population of massive perturbers, such as giant molecular clouds
or star clusters, that scatter efficiently field stars across the loss
cone boundary, leading to rapid coalescence of the binary$^{130,131}$.
In galaxies where none of these mechanisms is present and the binary
stalls, infall of a "third" massive black hole, during a repeated
merger, can accelerate the decay$^{133,134}$. A hierarchical triplet may form,
with the third black hole strongly perturbing the hard inner binary through
the exchange of energy and angular momentum as in a three-body
encounter. If
the orbit of the third black hole is inclined 
relative to the orbital plane of the inner binary, a Kozai resonance
lead the eccentricity of the inner binary to such an high value that
the two black holes coalesce promptly$^{135}$.

\bigskip
In the dense nuclei of the less massive ellipticals, two-body
relaxation may lead to the repopulation of the loss cone. Stars
diffuse across the loss cone boundary so that the binary decays on a
time $a/{\dot a}\propto t_{\rm rel}\propto N$.  This regime is
opposite to the collisionless case, and often ellipticals fall in
between, having relaxation times comparable to their age.  In this
case, gravitational encounters contribute to the refilling of the loss
cone but the phase-space distribution does not reach steady
state. Steep angular momentum gradients at the loss cone boundary are
produced after the sudden draining of stars, following the formation of
the hard binary. Since the collisional transport rate in phase-space
is proportional to these gradients, steep gradients imply an increased
flux of stars. This transitory enhancement becomes even most important
when the galaxy's nucleus experience repeated mergers or infall. In
this case the loss cone returns to an unrelaxed,
non-equilibrium state with a large angular momentum gradient$^{125}$.

\bigskip
>From the above considerations, refilling of the loss cone either
through non-equilibrium relaxation or dynamical relaxation implies a
depletion of stars {\it outside} the loss cone. This process destroys
the nuclear stellar cusp producing "missing light". This is at the hart
of core scouring by the black hole binary.  Since $M_{\rm ej}$ gives
the stellar mass removed by the binary, and since $M_{\rm ej}$
correlates with the binary mass, theory of binary decays finds its
most outstanding test in bright ellipticals built
from repeated mergers (as mass depletion requires at least ${\cal
{N}}M_{\rm ej}$ with ${\cal {N}}\sim  3-5$ mergers) where
black hole dynamics is dominated by the stellar component of the
interacting systems$^{92,93,120}.$

\bigskip
Real binaries likely evolve through a combination of different
mechanisms where gas plays also a role. There is a close parallelism
between the final parsec problem and the problem of quasar fueling:
both requires that a mass in gas and/or star comparable to $M_\bht$ or
even larger be supplied to the {\it inner parsec} of the galaxy host
within a timescale $\sim 10^{8-9}$ yr, much shorter than the age of
the universe.  Observations of powerful AGNs and quasars indicates
that Nature accomplishes this probably through inflows triggered by
torques from stellar bars$^{136}$ and/or unstable, massive gas
disks$^{137}$. There is the suspicion that the same inflow of gas that can
contribute to the fueling could contribute to the orbital decay of the
black hole binary in a number of ways: through viscous torques in
self-regulated accretion disks that extract angular momentum from the
orbit or through renewed star formation.

\section{Black hole pairing in gas-rich mergers}

The study of black hole dynamics from the state of pairing P down to
the GW domain in {\it gas-rich mergers} is a field in rapid progress.
The heavy computational demand and the rich physics involved require
to proceed through well defined steps and simplifications in order to
have full control of the basic physics and a strategy to attack the
complexity which is present when treating with star formation and
gas-dynamics.  The importance of black hole binary evolution in
gaseous background is not only limited to their dynamics. It also
involves the physics of accretion that determines the black hole mass
growth and spin evolution, and the interaction of the black holes with
the environment through their energy and momentum deposition in the
interstellar medium$^{8,10}$. These processes are inter-related and touch many
aspect of galaxy and AGN evolution. In describing the first key
results, it will become clear that the body of data from simulations
is still inhomogeneous, and many question remain unanswered.  While
black hole dynamics in collisionless systems has been explored mainly
in the hardening phase H where the numerical challenge is the highest,
black hole dynamics in dissipative/collisional systems has been mainly
studied from the onset of pairing, in gas-rich galaxies simulated with
some realism, down to the onset of the hardening phase H that occurs,
when/if present, inside a massive nuclear disk, immersed in a massive
bulge.  Trial runs with galaxies either gas-free or with a
hot-virialized gas component, have been carried out to test the
sensitive dependence of black hole pairing$^{138}$ on the background
models.  These trials reveal that we are still far from having a
unified view of the fate of black holes, and in the incoming
paragraphs we will summarize key findings and open problems.

\subsection{Rapid birth of a black hole binary in the aftermath of a major merger}

Galaxies, as observed in our low-redshift universe, are multi-component systems
of collisionless dark matter and stars that coexist in equilibrium with either a hot
or cold multi-phase gaseous component: a central massive black hole is also part of this landscape. 
Simulating a collision between two galaxies thus requires the versatile tool
of hydrodynamical simulations.

\medskip

The merger of two spiral galaxies similar to the Milky Way represents the prototype 
of a galaxy collision, and it has been studied in the literature typically with force
resolution of $\gsim 100$ pc$^{11,112,138-140}$. 
The merger of two spirals with black holes complements these studies, 
representing a real numerical challenge 
as in the simulation there is the need to 
reach/maintain a force resolution on a scale as small as $5-10$ pc in order 
to accurately trace the hole's dynamics down to $a^{\rm gas,star}_{\rm binary}$. 
This is the "minimal" request that can be attained with gradual
resizing of the computational volume via splitting techniques.

\medskip
Figure 7 shows the outcome of a merger, selected among a suite of
simulations with different encounter geometries and disk orientations
by Mayer et al.$^{141}$.  Here, twin Milky Way like galaxies approach
each other on a parabolic prograde coplanar orbit with pericentric
distance equal to $0.2$ of the galaxy's virial radius, typical of
cosmological mergers.  Each galaxy model comprises a central black
hole of $3\times 10^6\,\msun$ lying on the observed $M_\bh$ versus
$\sigma$ relation, a stellar bulge (modeled as an Hernquist sphere of
mass $M_{\rm Bulge}=0.008M_{\rm vir}$), an exponential disk of stars
and gas (with total mass $M_{\rm disk}=0.04M_{\rm vir}$ and gas
fraction $f_{\rm gas}=10\%$), and a massive ($M_{\rm
vir}=10^{12}\,\msun$) extended dark matter halo with NFW density
profile$^{142}$. Simulations are performed using the N-Body/SPH code
GASOLINE$^{143}$. Gas thermo-dynamics includes a recipe of star formation 
and radiative cooling down to a relatively high floor temperature of
20,000 K to account for turbulent heating, non-thermal pressure forces
and the presence of a warm interstellar medium.
 
 \bigskip
The galaxies first experience a few close fly-by (upper six panels of
Figure 7) during which tidal forces start to tear the galactic disks
apart, generating tidal tails and plumes.  The steep potentials at the
center of both galaxies allow the survival of their baryonic cusps
that sink under the action of dynamical friction against the dark
matter background, dragging together the two black holes.  During the
last close passage, prior merger, strong spiral patterns appear in
both the stellar and gaseous disks: non axisymmetric torques
redistribute angular momentum so that as much as 60$\%$ of the gas
originally present in each disk of the parent galaxy is funneled
inside the inner few hundred parsecs of the individual galaxy cores.
This is illustrated in the middle panel of the last row.  The black
holes, still in the pairing phase P, are found to be surrounded by a
rotating stellar and gaseous disk of mass $\sim 4\times 10^8\,\msun$
and size of a few hundred parsecs. The two black holes with their own
disks are just 6 kpc far apart and accretion (not resolved on this
scale) is expected to light up the two black holes as in the case of
NGC 6240. In the meantime a star-burst of 30 $\msun$ yr$^{-1}$ has
invested the central region of the on-going merger. It is worth noting
that already at this stage the dissipative role of gas changes the
black hole environment. While in the collisionless case (as depicted
in Figure 4) the black holes pair inside their stellar
cusps that preserve their density, in the dissipative case they pair
inside a denser background resulting from gas-dynamical instabilities
that have funneled additional mass in the hole's surroundings.

\begin{figure}
\centering{
\resizebox{7.5cm}{!}{\includegraphics{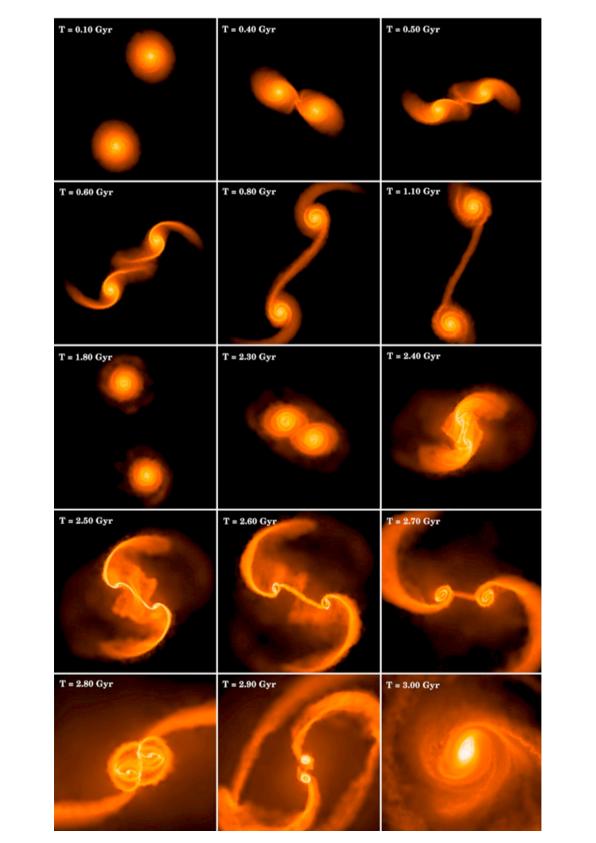}}\\
\caption{A major merger between two Milky Way like galaxies$^{141}$:
the simulation follows the evolution of dark matter, stars, gas, and
of the massive black holes ($M_\bh=3\times 10^{-3}M_{\rm Bulge}$), but
only the gas component is visualized for seek of clarity. Brighter
colors indicate regions of higher gas density and the time
corresponding to each snapshot is given by the labels. The first 10
images measure 100 kpc on a side, roughly five times the diameter of
the visible part of the Milky Way galaxy. The next five panels
represent successive zooms on the central region. The final frame
shows the inner 300 pc of the nuclear region at the end of the
simulation.} }
\label{fig:ngc3341}
\end{figure}
\bigskip

At this stage of the simulation, a computational volume of 30 kpc in
size is refined in order to achieve a force resolution of $2$ pc and
continue the study of the black hole's dynamics.  The fine-grained
region selected is large enough to guarantee that stars and dark
matter particles essentially provide a smooth background, while the
computation focuses on the gas component which dominates by mass in
the region.  The short dynamical timescale involved in this phase
($1-10$ Myr) compared to the star-burst duration ($\sim 100$ Myr)
suggests to model the thermodynamics and radiation physics simply via
an effective equation of state.  Calculations that include radiative
transfer show that the thermodynamic state of a metal rich gas heated
by a star-burst, comprising stellar feed-back, can be approximated by
an equation of state of the form $P=(\gamma - 1) \rho u\,$ with
$\gamma=7/5$ where the specific internal energy $u$ evolves
with time as a result of $PdV$ work and shock heating$^{144,145}$.

\bigskip
With time, the two baryonic disks eventually merge in {\it a single
massive self-gravitating, rotationally supported nuclear disc} of
$\lsim 100$ pc in size, now weighing $3\times 10^9\,\msun$.  This {\it
grand disk} is more massive than the sum of the two nuclear disks
formed earlier because radial gas inflows occur in the last stage of
the galaxy collision.  A strong spiral pattern in the disk produces
remarkable radial velocities whose amplitude declines as the spiral
arms weaken over time.  Just after the merger, radial motions reach
amplitudes of $\sim 100\, \kms$, and possibly feed the black holes on
flight.  The nuclear disk is surrounded by a more diffuse,
rotationally-supported envelope extending out to more than a $\sim$
kpc from the center and by a background of dark matter and stars
distributed in a spheroid.  This grand disk is rotationally supported
($v_{\rm rot} \sim 300$ km s$^{-1}$) and also highly turbulent, having
a typical velocity $v_{\rm turb} \sim 100$ km s$^{-1}$.  Multiple
shocks generated as the galaxy cores merge are the main source of this
turbulence.  The disk is composed by a very dense, compact region of
size about 25 pc which contains half of its mass (the mean density
inside this region is $> 10^5$ atoms cm$^{-3}$). The outer region
instead, from 25 to 75-80 pc, has a density 10-100 times lower.  The
disk scale height also increases from inside out, ranging from 20 pc
to nearly 40 pc.  The volume-weighted density within 100 pc is in the
range $10^3-10^4$ atoms cm$^{-3}$ compatible to the observed nuclear
disks$^{146-149}$. This suggests that the degree of dissipation
implied by the effective equation of state gives a reasonable
description of reality despite the simplicity of the thermodynamical
scheme adopted.

\bigskip
After 5.12 Gyr from the onset of the collision, the black holes,
dragged toward the dynamical center of the merger remnant, move inside
the grand disk spiraling inward under the action of gas-dynamical
friction.  In less than $\lsim 1$ Myr, a timescale much shorter than
the duration of the star-burst, the relative black hole orbit decay
from $\sim 100$ pc to $\lsim 5$ pc where they form a {\it binary}, as
mass of gas enclosed within their separation is less than the mass of
the binary. It is the gas that controls the orbital decay, not the
stars.  The transition between state P to B is now completed as
illustrated in Figure~8: a mildly eccentric binary ($e\sim 0.5$) has
formed with orbital period of $10^5$ yr.  This short sinking timescale
comes from the combination of two facts: that gas densities are higher
than stellar densities in the center due to the dissipative nature of
the interaction, and that the black holes move relative to the
background with mildly supersonic velocities where the drag is the
highest$^{141}$.  But, how sensitive are the results on the input
physics?

\bigskip

Mayer et al.$^{141}$ have decreased the degree of dissipation by
increasing $\gamma$ to $5/3$, to mimic the presence of a heat source.  Is is found
that in this warmer environment, a turbulent, pressure supported cloud
of a few hundred parsecs forms rather than a rotationally supported
cold disk. The mass of gas is lower within 100 pc relative to the
$\gamma=7/5$ case because of adiabatic expansion following the final
shocks, at the time of merging of the two cores. The nuclear region is
still gas dominated, but the stars/gas ratio is $> 0.5$ in the inner
$100$ pc.  The black holes in this hotter environment follow a
"retarded" dynamics: the transit from P $\to$ B is not any longer seen, 
over a time span of $\lsim 1 $ Myr.
The black holes remain at a distance of 50-150 pc, as
shown in Figure 8, and a bound system is expected to form, according
to eq. 9, on a time scale $t_\df\sim 10^7$ yr, under the
joint control of stars and of the lower density gas.

\begin{figure}
\centering{
\resizebox{8cm}{!}{\includegraphics{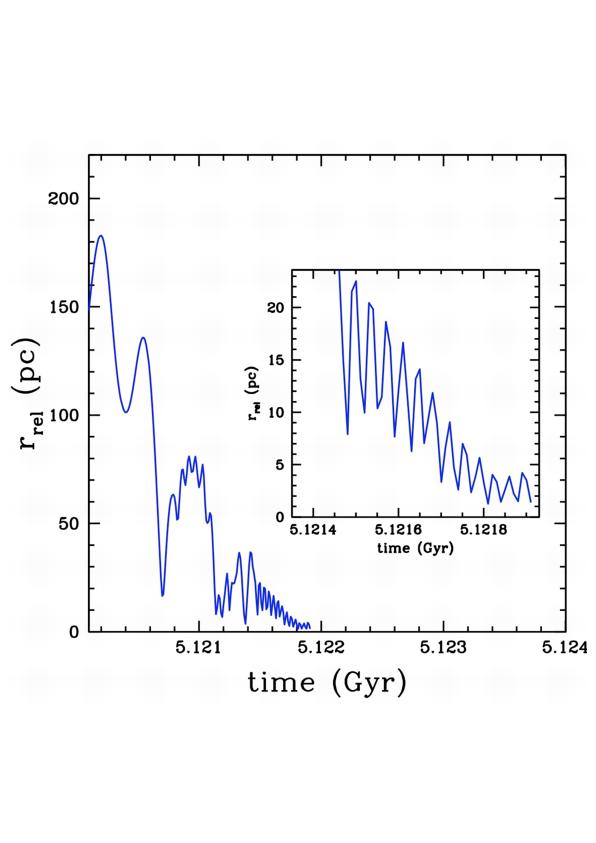}}
\resizebox{8cm}{!}{\includegraphics{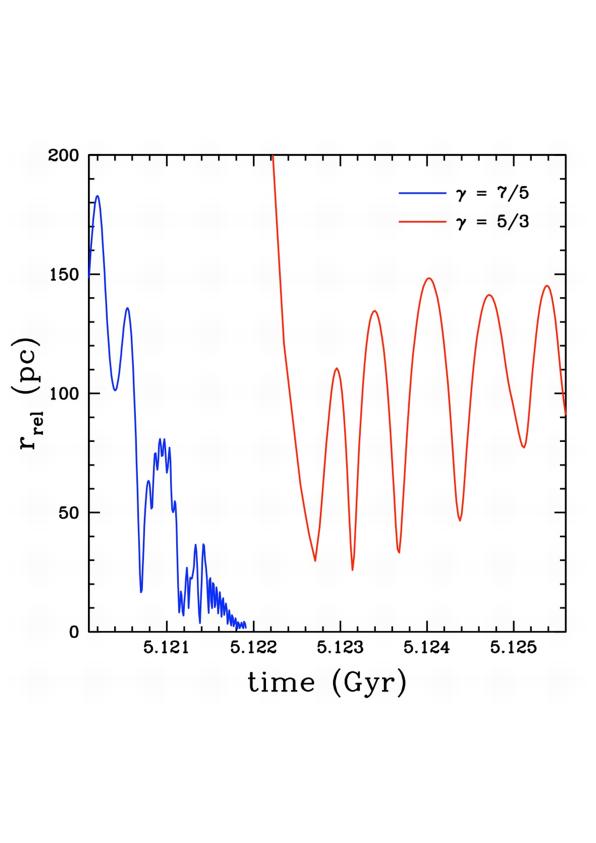}}\\

\caption{ Orbital separation of the two black holes as a function of
  time during the last stage of a selected galaxy merger when a
  massive nuclear disk forms$^{141}$. The orbit of the pair is eccentric until
  the end of the simulation. The two peaks at scales of tens of
  parsecs mark the end of the phase during which the two holes are
  still embedded in two distinct gaseous cores. Until this point the
  orbit is the result of the relative motion of the cores combined
  with the relative motion of each black hole relative to the
  surrounding core, explaining the presence of more than one orbital
  frequency.  The inset shows the details of the last part of the
  orbital evolution which  takes place in the nuclear disc arising from
  the merger of the two cores. The binary stops shrinking when the
  separation approaches the softening length (2 pc).  In the right
  panel, the blue line shows the relative black hole distance as a
  function of time for the run with $\gamma=7/5$ as shown in the left
  panel, while the red line shows it for $\gamma=5/3.$ }}
\label{fig:pairing}
\end{figure}

\bigskip
The radiative injection of energy from the black holes accreting on
flight might be a candidate for a heating source, and for increasing
the turbulence in the gas by injecting kinetic energy$^{150,151}$ in
the surrounding medium, possibly creating outflows.  A rough estimate
for the energy released in the accretion process is $\Delta E_{\rm
acc}\sim f_{\rm dep} (L/ L_E) (t/t_{\rm Ed}) \epsilon M_{\rm BH}c^2 $
(with $L_E$ the Eddington luminosity, $\epsilon\sim 0.1$, and $f_{\rm
dep}$ the fractional energy deposited by the accreting black hole in
the surrounding gas).  This energy is $\Delta E_{\rm acc}\sim  2\times
10^{57} \rm {erg}$, for a black hole mass of $10^6\,\msun$ accreting at
the Eddington limit for a time $t_{\rm Ed}\sim 10^7$ yr, and $f_{\rm
dep}= 0.01$.
This energy appears to be larger than the energy dissipated by
the binary in its inspiral down to $a_\gw$, 
$\Delta E_{\rm B}=GM_\bht \mu_\bh/2 a_\gw
\sim 6\times 10^{54}\,\rm {erg}$.
Energy released on flight by the black holes
can influence the thermal state of the grand disk which is
supposed to convert a sizable fraction of its gas into stars.
This environmental change is expected to affect the dynamics 
and the fueling of black holes (even after black hole coalescence) 
with consequences that have not been exploited yet, given the
computational challenge that these processes pose.

\bigskip 
Increasing the resolution will be the next priority.  At higher
spatial resolutions ($\gsim 10^{-3}$ pc),  the grand disk is expected to
reveal a much richer structure with spiral patterns and bars,
considered as possible candidates for triggering small scale inflows
around the black holes.  Regions that appear featureless in low
resolution runs should instead reveal the true nature of the gas.
Wada \& Norman$^{152}$ showed that 
when the main radiative processes and the
feedback from supernovae are directly incorporated in simulations of
rotating gaseous disks, a multi-phase turbulent interstellar medium
arises naturally due to the interplay between gravitational/thermal
instabilities and energy injection by supernovae. The resulting
structure is clumpy at all scales.  This is a clear example that
there is still some {\it missing} physics that needs to be
incorporated, beyond polytropes.

\subsection{Pairing in minor mergers}

Pairing seems inescapable in major, equal-mass mergers: the violence
of the collision affecting both galaxies in a nearly symmetric fashion
(regardless the geometry of the encounter) brings the two black holes
in the central region of the remnant. There, in presence of a cold
nuclear disk, a binary forms rapidly within a Myr after the merger is
completed.  But, what about pairing in minor merger?  Minor,
i.e. unequal-mass mergers differ profoundly from major mergers.
Encounters are uneven, resembling more an "accretion" process than a
merger as the less massive galaxy is dramatically damaged during its
sinking into the primary by tidal and ram pressure forces.

\bigskip

Simulations by Boylan-Kolchin$^{153-155}$ of collisionless spherical halos
shows how details on the encounter geometry, satellite-to-host mass
ratio, internal structure (dark matter plus bulge), and on presence
(absence) of a central black hole in the main host affect the degree
of damage of the satellite resulting in a broad variety of outcomes.
Satellites on less-bound, eccentric orbits are more easily destroyed
than those on more-bound or more-circular orbits as a result of an
increased number of pericentric passages and greater cumulative
effects of gravitational shocking and tidal stripping. In addition,
satellites with densities typical of faint elliptical galaxies are
disrupted relatively easily, while denser satellites can survive much
better in the tidal field of the host.  Also satellites merging on to
a host with a central black hole are more strongly disrupted than
those merging on to hosts without the black hole.  Thus, it is not
surprising that also black hole pairing is affected in major ways by
these details. It is clear that investigating the necessary conditions
for pair formation in minor mergers with gas is of primary importance,
both in a broad cosmological context,and for the impact on the {\it
LISA} data stream.

\bigskip

Callegari et al.$^{156}$ find different outcomes, in their systematic
study of 4:1 (i.e.  mass ratio $q=0.25$) and 10:1 (i.e. $q=0.1$)
collisionless and gas-dynamical mergers, performed with GASOLINE at
various cosmic epochs.  Specifically, the 4:1 mergers are studied in
the local Universe ($z=0$), while the 10:1 mergers at $z=3$, assuming
a constant $M_{\rm BH}-M_{\rm Bulge}$ relation in between these cosmic
epochs$^{4}$, and rescaled models, replica of the Milky Way, for the
galaxies. In the $z=3$ runs, the black hole masses are thus
$6\times10^5$ and $6\times10^4\,\msun$, so that their expected
inspiral and coalescence signal falls nicely in the {\it LISA}
sensitivity window$^{29,30}$.

\begin{figure} 
\begin{center}
\includegraphics[width=0.50\textwidth]{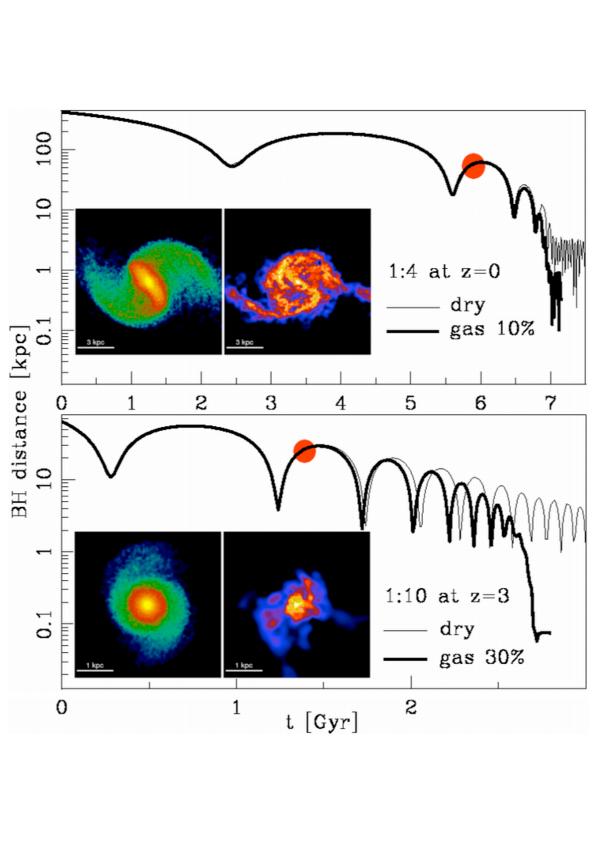}
\end{center}
\caption{ Black hole relative separation as a function of time in four
simulations$^{156}$.  In the upper row, the hole's distance refers to
two selected 4:1 mergers (for galaxy models at $z=0$); the thin and
thick lines correspond to simulations with no-gas (dry) and with gas
($f_{\rm gas}=0.1\%$), respectively.  In the lower row, the hole's
distance refers to 10:1 mergers (for galaxy models at $z=3$); the thin
and thick lines correspond to simulations with no-gas (dry) and with
gas ($f_{\rm gas}=0.3\%$), respectively.  The insets show the
color-coded density maps of stars (left) and gas (right), on a box of
4 kpc on a side.  The large dot on each curve indicates the time at
which the two snapshots are recorded.  Colors code the range
$10^{-2}-1$~M$_\odot$~pc$^{-3}$ for stars, and
$10^{-3}-0^{-1}$~M$_\odot$~pc$^{-3}$ for the gas.  These snapshots are
representative of the average behavior of the satellites during the
first two orbits.  Note the formation of a strong bar for the 4:1
minor merger, which is absent for the 10:1 case, and the truncation of
the gaseous disk in the 10:1 satellite caused by ram pressure
stripping.}  
  \label{fig:inflow}
\end{figure}

\bigskip

All collisionless cases, regardless the value of $q$, 
do not lead to the formation of a close black hole pair: tidal shocks progressively
lower the density in the satellite until it dissolves, leaving a
wandering black hole in the remnant.  Only with the inclusion of
a gaseous disk component,  inclusive of radiative 
cooling, star formation, and supernova feedback, 
the merger changes significantly. Figure 9 depicts the   
star-forming gaseous component of the satellite, to show
its profound structural damage (the primary is not shown). 
For mass ratios 4:1 at $z=0$, bar instabilities excited at pericentric
passages funnel gas (present in a fraction $f_{\rm gas}=0.1$ of the total
disk mass) to the center of the satellite, steepening its potential
well and allowing its survival against tidal disruption down to the
center of the merger remnant.  Therefore in this case, the presence of
a dissipative component is necessary and sufficient to pair the black
holes down to the force resolution limit of $\sim100$~pc scales, and
to create conditions favorable to the formation of a binary.  Instead,
the smaller satellites (in the 10:1 mergers at $z=3$) are more
strongly affected by both internal star formation and the
gas-dynamical interaction between their interstellar medium and that of the primary
galaxy. Torques in the early stages of the merger are not acting to
concentrate gas to the center, due to the absence of a stellar bar and
the stabilizing effect of turbulence. As a result, ram pressure strips
all of the interstellar medium of the satellite. Still, gas-rich
satellites (with $f_{\rm gas}=0.3\%$) undergo a central burst of star
formation during the first orbits which increases their central
stellar density, thus allowing their survival and ensuring the pairing
of the two black holes via dynamical friction in a few $10^8$~yr after
the disruption of the satellite.  By contrast, in a satellite with
lower gas content ($f_{\rm gas}=0.1\%$) star formation is not effective
in enhancing its concentration: as a consequence, even its central
region is disrupted at a few hundred parsecs from the center of the
primary, and the pairing between the two black holes is delayed by a
few billion years.  The sensitive dependence of pairing in minor
mergers suggests that there exist a {\it minimum} mass ratio for the
interacting galaxies below which phase B is unattended. These
simulations$^{156}$ suggest that a 10:1 merger is nearing this limit,
depending sensitively on the fraction of gas present in both galaxies.

\section{Black hole dynamics in self-gravitating nuclear disks}

As shown in Section 4.1, major mergers between multi-component
galaxies lead to the pairing of black holes inside the turbulent,
gaseous disk that forms in the aftermath of the collision.  Ideally, a
large set of simulations should be carried on to explore how black
holes transit from P$\to$B$\to$H, under a variety of initial
conditions.  To encompass the enormous computational effort, a number
of authors explored (using GADGET) black hole sinking in a two-composite
system, i.e. an equilibrium circum-nuclear disk embedded in a stellar
bulge$^{157-131}$, viewed as relics of the merger.  In all these
studies, the equilibrium disk is modeled as a self-gravitating Mestel
disk of size ($R_0\lsim 100$ pc), mass ($M_{\rm disk}=10^8\,\msun$)
and vertical scale height ($H\sim 10$ pc), as close as indicated by
observations of massive nuclear disks in Seyfert and
ULIRGs$^{117-120}$.  Stars follow a Plummer profile and their mass
within the spherical volume of radius $R_0$ is five times the disk
mass, as suggested by observations$^{146}$.

\bigskip

In modeling the gas disk, thermodynamics plays a critical role.  In
real astrophysical disks, massive gas clouds coexist with
warmer phases and a simple isothermal equation of state would provide
an unrealistic averaged representation of the real thermodynamical
state.  Cold self-gravitating disk are unstable to fragmentation and
find their stability as soon as stars, resulting from the collapse
and/or collision of clouds, inject energy in the form of winds and
supernova blast waves, feeding back energy into the disk now composed
of stars and a multiphase gas.  Due to the complexity of implementing
in a code this rich physics at the required level of accuracy, a
simple polytropic equation of state of the form $P=K\rho^\gamma$ is
introduced to capture the basic thermodynamical behavior of the gas.
$K$ is a free parameter corresponding to the entropy of the gas and
$\gamma$, varying between $1$ and $5/3$, is a parameter that can be
tuned to represent either a pure adiabatic response ($\gamma=5/3$) or
a response that mimics cooling, star formation and feedback
($\gamma=7/5$) as in Mayer et al.$^{141}$.

\bigskip
Escala et al.$^{128}$ consider cases where $K$ varies to reproduce
different degree of clumpiness, keeping $\gamma$ fixed to 5/3, while
Dotti et al.$^{131}$ explore, in a complementary fashion, cases with
$\gamma=5/3$ and 7/5, and fixed $K$.  Escala et al.$^{157}$ assume
equal mass black holes moving on circular orbits, varying the black
hole mass to disk ratio, and the angle between the orbital plane of
the pair and the circum-nuclear disk. Dotti et al.$^{158-160}$ explore
the dynamics of both equal and unequal mass black holes starting from
eccentric, prograde or retrograde orbits that may arise from arbitrary
encounter geometries and minor merger events.

\bigskip

The main results of all these works can be summarized as follows: 
(i) the black holes always reach  separations of 
a few parsecs in $\lsim 10$ Myr, and form a Keplerian binary;
(ii) the pairing process is faster in colder disks and for
black holes orbiting in the disk plane; 
(iii) black holes on initially eccentric prograde orbits lose memory of their 
initial eccentricity and form binaries consistent with $e\sim 0$; 
(iv) retrograde eccentric orbits
turn into prograde near-circular orbits before the black holes form a bound system;
(v) very massive binaries perturb the disk globally
and  excavate  transitory gaps in the circum-binary environment.

\subsection{Black hole inspiral, circularization and orbital angular \\ momentum flip}
\begin{figure} 
\centering{
\resizebox{10.cm}{!}{\includegraphics{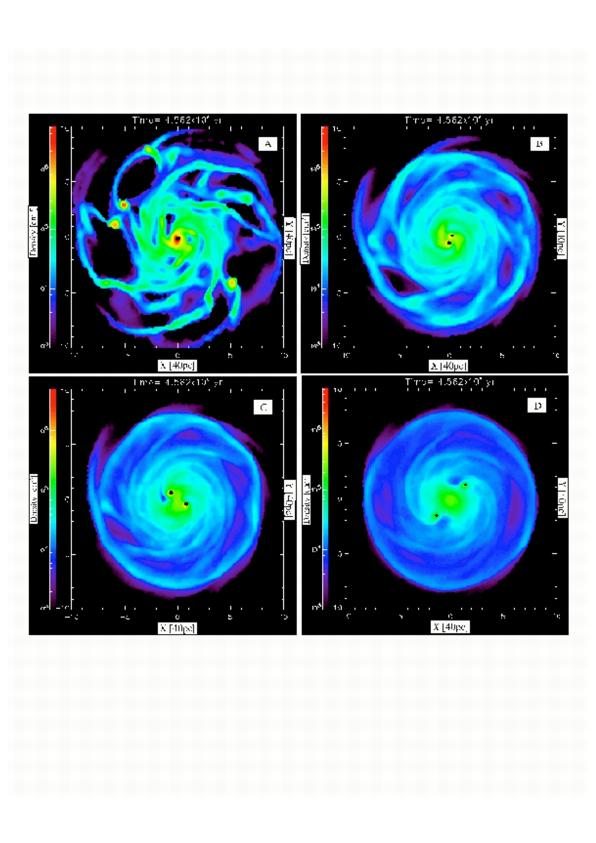}}
\caption{Face on view of the density distribution of a gaseous Mestel
disk hosting two equal mass black holes$^{157}$ (with $M_\bh=0.01
M_{\rm disk}$).  The black dots indicate their position, recorded at
the same time.  The disk clumpiness varies according to the parameter
$K$.  The form of the density distribution is smooth for large values
of $K$ (run D) and shows spiral waves, while it is highly
inhomogeneous for small $K$, as dense clumps and filaments form as a
result of a lower pressure support against gravitational pull.  }  }
\label{fig:miewakes}
\end{figure}
In the rotating background of a Mestel disk, the black holes on
circular orbits excite prominent wakes both in front of and behind
their path, but the wake that forms behind is the most effective in
exchanging angular momentum given the rotation pattern and surface
density profile of the disk$^{158}$.  This causes a net braking and the black
holes end forming a binary in less than 10 Myr.

\bigskip

Braking by gas-dynamical friction depends also on the level of disk
clumpiness that introduces a noise element on the black hole dynamics:
clumps perturb the black hole orbits more strongly and at random
phases so that their inspiral is more rapid and chaotic.  In smoother
background the perturbation instead takes the form of a spiral wave.
Figure 10 depicts snapshots of the face-on density distribution of the
gas disk and current black hole position, obtained for four different
runs with increasing $K$, and computed at the same time, from 
Escala et al. $^{157}$.  In the cold
clumpy disk, black holes have decreased their separation by more than
a factor 10 reaching, the numerical resolution length.  The
sensitivity to clumpiness suggests a parallelism with stellar systems:
as large clumps (likely in the form of dense molecular clouds)
accelerate binary hardening in gaseous disks, the same holds true in
stellar backgrounds where molecular clouds accelerate loss cone
refilling and so black hole inspiral.

\bigskip
Circularization in a smooth (high $K$) rotating disk is also a clear
prediction of these studies$^{159}$.  When the black holes move on eccentric
orbits, circularization takes place well before the black holes form a
binary and this is a feature peculiar to dynamical friction in a
rotating background.  On an initially eccentric orbit, the black hole
has, near periapsis, a velocity larger than the local rotational
velocity, so that dynamical friction causes a reduction of its
velocity: the wake of fluid particles lags behind the black hole
trail.  By contrast, near apoapsis, the black hole velocity is lower
than the disk rotational velocity, and in this case, the wake is
dragged {\it in front} of the black hole's trail, causing a positive
tangential acceleration. The composite effect is a decrease of $e\to
0$.  In spherical backgrounds, dynamical friction tends to
increase/maintain the eccentricity, both in collisionless$^{116, 161,
162}$ and in gaseous$^{163}$ environments.  This is suggestive that
the presence of a rotating gaseous background can be inferred from the
smallness of the binary eccentricity on scales of a few parsecs.

\bigskip

A further recently discovered signature of a rotating background is
the angular momentum flip of initially counter-rotating
orbits$^{160}$.  This has been highlighted when considering the
inspiral of a black hole (called secondary) moving in a Mestel disk
with $\gamma=7/5$ and hosting at the center a primary black hole. The
initially negative angular momentum of the moving black hole grows
very fast during the first million year.  The black hole-disk
interaction brakes the hole at all orbital phases, because the hole's
velocity and disk flow pattern are anti-aligned and the density wake
lags alway behind the hole.  The increase of the orbital angular
momentum is further facilitated by the fact that, while the orbit
decays, the black hole interacts with progressively denser regions of
the disk. The orbit is nearly radial at the time the orbital angular
momentum changes sign and this occurs before the secondary binds to
the central black hole. When co-rotation establishes, the orbital
momentum increases under the circularizing action of dynamical
friction in its co-rotating mode.  This orbital {\it flip} has been
visible thanks to the high numerical resolution adopted (0.1 pc
through the entire simulation) that allows to trace the gas-black hole
interaction with unprecedented accuracy.  In conclusion, a further
prediction of black hole inspiral in rotating disks in that
gas-dynamical friction conspires to drive counter-rotating orbits in
co-rotation before the time of formation of a Keplerian binary.

\subsection {Accretion on flight}
\bigskip
Accretion can be studied only if the resolution length-scale of the
code is smaller than the Bondi-Hoyle-Littleton radius of the moving
black hole.  If this resolution is achieved, as in a recent series of
simulations$^{160}$, its is possible to follow the fueling of the
black hole {\it on flight}, i.e along its motion, and also to include
the contribution of the drag force ${\bf F}^{\rm gas}_{\rm BHL}$ due
to accretion from the numerical self-consistent measure of $\dot M.$
While in previous studies, each black hole was treated as a
collisionless particle, now the hole is a {\it sinking} particle,
ready to capture gas within $R_{\rm BHL}$ while orbiting inside the
disk.  

\bigskip

In a recent study$^{160}$, accretion onto the secondary, moving black
hole has been traced starting either from a prograde or a retrograde
eccentric orbit relative to the disk's rotation.  It is found that as
long as the orbit of the moving black hole remains eccentric,
accretion occurs at a low rate, with an Eddington factor $f_{\rm Ed}$
(defined as the mass accretion rate $\dot M$ in units of the Eddington
rate) close to 0.1, due to the large relative velocity between the black hole
and the fluid, and is highly variable owing to the presence of
inhomogeneities in the gas in the vicinity $R_{\rm BHL}$.  Only when
the orbit has circularized (and flipped for the retrograde case), the
moving black hole accretes at high rate with $f_{\rm Ed} \sim 0.9$ and
in a less variable fashion.  Thus, two accretion phases are present
for the moving black hole: the first is the low/variable state that
occurs before circularization (and/or angular momentum flip) and the
second is the high/constant state after circularization in the
co-rotating mode.  By contrast, the central black hole undergoes
accretion at an almost constant rate $f_{\rm Ed} \sim  0.9$, in the
absence of a substantial motion.

\bigskip
What is the fate of gas inside the black hole sphere of influence
$R_{\rm BHL}$?  In the simulations, it is possible to keep record of
the specific energy and angular momentum of every accreted fluid particle
as a function of time.
As the gas density in the black hole vicinity (near $R_{\rm BHL}$)
can be as high as $10^7$ cm$^{-3}$, it is 
conceivable that dissipative/radiative processes operate to reduce the gas
internal energy below the values adopted in the simulations.  In these
circumstances, we expect that the fluid particles bound to each black
hole form a geometrically thin, accretion disk in Keplerian rotation.
Since every gas particles is found to carry a net angular momentum in the
direction parallel to that of the grand nuclear disk, the size of the
coplanar accretion disk can be computed according to angular momentum
conservation: $R_{\rm BH,disk}\sim L^2_{\rm ver}/(G\,M_\bh)$, where
$L_{\rm ver}$ is the vertical component of the specific angular momentum of the
particles bound, relative to each hole.  After circularization,
$R_{\bh, \rm disk}$ can be as small as  $0.01$ pc for the very bound particles, 
strongly suggesting that both black holes
are fed through a thin, cold accretion disk along the course of
the inspiral$^{159}$.  Viscosity (of magneto-rotational origin) is expected to
self-regulate the transport of angular momentum in the small Keplerian
disks, so that gas flows onto each black hole at the rate determined
by the physical conditions of the gas at the outer boundary.

\begin{figure} 
\centering{
\resizebox{8cm}{!}{\includegraphics{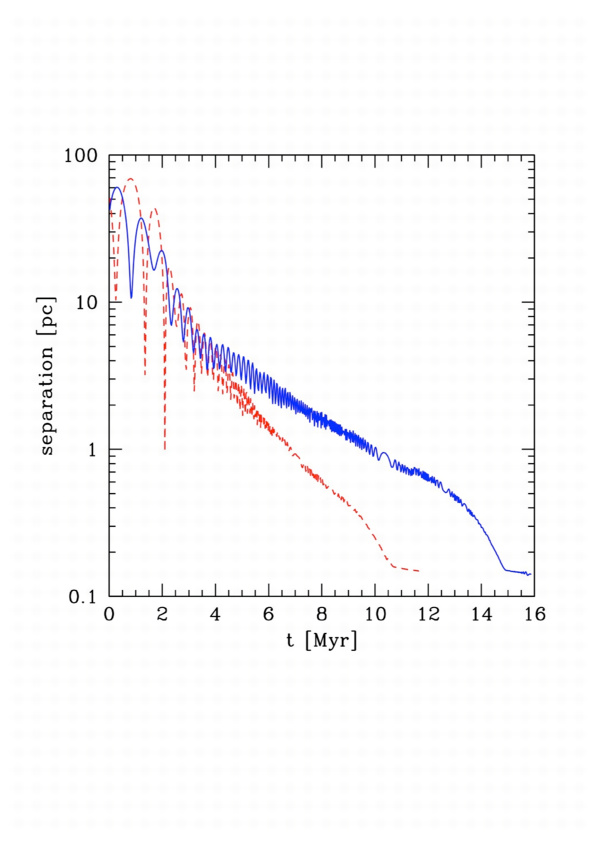}}
\caption{Black hole relative separation as a function of time, as
inferred from two high resolution simulations by Dotti et
al. (in preparation). Blue-solid (red-dashed) line refers to the case where the
moving black hole is set initially on a co-rotating (counter-rotating)
orbit with $e=0.7$.  In the gaseous background of the rotating Mestel
disk (with $\gamma=7/5$) the binary decays transiting across the three
phases, from pairing P, to B where a Keplerian binary forms, down to
the hardening phase H.  }}
\label{fig:highres}
\end{figure}

\subsection{Hardening in a gaseous environment: binary stalling \\
on sub-parsec scales? }

With increased {\it spatial} and {\it mass} numerical resolution,
binary hardening has now being explored down to separations close to
$R_{\rm BHL}$.  Figure 11 illustrates the outcome of the high
resolution simulation by Dotti et al. 
describing the hardening process through phases
P $\to$ B $\to$ H, for a prograde and retrograde initially eccentric
orbit.  Orbital decay, guided by gas-dynamical friction, continues
from the $\sim 10$ pc scale down to $\sim 0.5$ pc, maintaining the same
derivative across this domain.  When the black hole separation is
$\lsim 0.5$ pc the derivative steepens indicating that the nature of
the black hole-disk interaction has changed.  As first indicated by
Escala et al.$^{124}$, torques resulting from the ellipsoidal deformation,
arising from the superposition of the interacting wakes, accelerates
the orbital decay.  Only when the black hole separation reaches the
resolution limits of the gravitational and hydrodynamical forces the
binary stalls, as artifact of the numerical resolution.  The rapid
decay is a feature present in the $\gamma=7/5$ runs, i.e. when the
thermodynamical treatment allows from some degree of
dissipation. Equivalent runs for $\gamma=5/3$ show evidence of
stalling suggesting that dissipative processes play a critical role in
determining the black hole dynamics and fate below $\sim 1$ pc.

\bigskip

Reaching the scale $a_\gw$ 
is still a numerical challenge 
because not only it requires an increasingly
high resolution, but new input physics.
The physics that is described in the next section will relay
on the assumption 
that the gaseous nuclear disk remains on all scales gaseous, i.e. immune by episodes 
of star formation and feed-back, and that in the near vicinity of
the black hole binary the gas has cooled down to
form a {\it circum-binary} disk.
The black holes them-selves carry their own small,  tidally truncated
accretion disks (confined within their Roche lobes).
These disks  can be either consumed or replenished, but their presence 
will be ignored in the next Section as they play a marginal role
in the hole's dynamics. Only in Section 6, these disks
will be considered as they play a major role
in determining the electro-magnetic signature of binary black holes
during their inspiral down to coalescence, and after their merger.

\subsection{Circum-binary discs and black hole migration}

Consider the case of two equal-mass black holes surrounded by a
circum-binary disk.  What is the physical state of this disk?  At the
radii of greatest interest, typically of $0.1$ pc, the circum-binary
disk is expected to be {\it self-gravitating} as it likely carries a
mass $M_{\rm CBD}$ in excess of $\sim 0.1 M_\bht$: self-gravity
controls the gas dynamics so that the circum-binary disk differs from a
canonical, geometrically thin accretion disk$^{164}$.  In this {\it
hybrid} configuration, the rotation curve is not fully Keplerian
(with $\Omega_{\rm K}=(GM_\bht/R^{3})^{1/2}$), the
surface mass-density $\Sigma(R)$ still declines as in a Mestel disk
($\Sigma\propto R^{-1}$), and the vertical scale height, which for a
Keplerian disk is $H_{\rm K}\sim c_{\rm s}/\Omega_{\rm K}$, takes instead  the
expression $H\sim c^2_{\rm s}/\pi G\Sigma$.
 
 \bigskip 
The stability of self-gravitating accretion disks around {\it single}
black holes has been investigated extensively in recent
years$^{165-169}$.  In these disks, large scale disturbances excited
by self-gravity cannot always be stabilized nor by pressure gradients
(at short wavelengths) nor by rotation (at longer wavelengths).  When
the non-linear outcome of the instability is particularly violent,
fragmentation of the disk into gravitationally bound clumps occurs,
turning gas into stars. But there are regimes of interest here, where
self-gravity affects the disk structure by exciting non-axisymmetric
instabilities that redistribute the angular momentum across the disk
without major damage.  Crudely, the dividing line from these two
different behaviors is set by the Toomre parameter $Q={c_{\rm s}\Omega/ \pi
G \Sigma} ,$ varying with distance $R$ across the disk: disks are
unstable to fragmentation in regions where $Q<1;$ those with higher
sound or turbulence speeds and lower surface densities tend
instead to be stable and have $Q\gsim 1$ at all radii.

\bigskip
Astrophysical disks may form 
initially with $Q\gsim 3$, but radiative losses are expected to cool the disks down, 
until eventually $Q\lsim 1.$   Numerical simulations show that
at this stage, any perturbed disk develops
a gravitational instability in the form of a spiral pattern. Compression and shocks 
induced by the instability lead to a net energy redistribution/deposition and 
the disk heats up.  The instability thus self-regulates its growth 
in a way that the disk is kept close to marginal stability.
Thus astrophysical disks are expected to evolve into a 
quasi-stable stationary state down to  $Q\sim  1$, and to
sustain  an {\it accretion} flow with outward angular momentum 
transport  through spiral waves$^{169}$.  

\bigskip

Circum-binary disks around massive black hole binaries are poorly known
theoretically. We do not know 
how they are fed from the outer massive
nuclear disk present on the $\sim 100$ pc scale, and how they evolve.
When a black hole binary is at the center of 
a self-regulated disk, its time-varying gravitational
field is expected to perturbe the disk, and in particular its inner
rim.  The binary, acting as a source of angular momentum, exerts a
tidal torque that inhibits the gas from drifting inside its
orbit. This creates a hollow density region, called {\it
gap}$^{170-174}$, that surrounds the binary. Further interaction
between the binary and the disk depends on the way the excess
of orbital angular momentum of the binary, transmitted to the disk, is
redistributed through viscous torques and/or spiral
patterns in the fluid, on the way the disk is fed and on its stability against
fragmentation.  

\bigskip
Gap formation can be understood, in these disks, using dynamical
friction as key concept.  Consider, for simplicity, the only presence
of an inviscid ring of gas skimming the black holes on a circular orbit at a
radius $R$ slightly larger than the binary separation $a$ ($R\gsim
a$), and focus on the fate of a fluid parcel. As the fluid parcel 
moves more slowly than the black holes, it is exerting a drag that tends to
decelerate the black holes.  The fluid element, accelerated
tangentially by the gravitational pull of the black hole closest to
it, reacts drifting back with an excess of orbital angular
momentum. As a result the entre ring will eventually drift away increasing its
distance from the binary: the annular region that was filling the
space near $a$ moves away increasing the gap size.  If a
viscous (continually fed) accretion  
disk is present outside the binary in place of the annular ring, the
response of the disk is determined by the competition between the
tidal torque exerted by the binary and the pressure and/or viscous
torques present inside the disk that, in opposition to the tide, tend
to refill the emptied region and keep the gap closed.  A hollow, low density
region (the gap) of size $\Delta\gsim a,$ surrounding the binary,
forms as long as $\Delta$ is larger than the Roche lobe radii,
i.e. the radii of the two equipotential surfaces connecting the two
black holes ($\Delta>R_{\rm RL}$), and the disk scale height
$H<\Delta$.  The gap is maintained when the gap-closing time $\Delta
t_{\rm close}$, comparable in magnitude to the viscous/turbulence time
at the inner edge of the disk $\sim t_{\rm vis/turb} (a)\sim
\Delta/v_{\rm drift/turb}$ (where $v_{\rm drift/turb}$ is the radial
drift/turbulent velocity of the disk near $a$) is longer than the
gap-opening time $\Delta t_{\rm open}\sim \Delta L/T$, where $\Delta
L\sim  (\rho_{\rm gas} H \Delta ^2 )RV_{\rm cir}$ is the angular
momentum that must be added to the gas to open a gap of size $\Delta$,
and $T$ is the tidal torque by dynamical friction that can be
estimated as $T\sim R F^{\rm gas}_{\rm DF}\sim R\,4\pi \ln \Lambda
\rho_{\rm gas} (GM_{\rm \bht}/V_{\rm cir})^2$ (from eq. 10).  The
condition $\Delta t_{\rm close}>\Delta t_{\rm open}$ is fulfilled in
presence of a massive binary, i.e.  whenever the gas mass enclosed
within $\sim a$ (estimated before gap opening) $M_{\rm gas}(a)
\lsim 0.5M_\bht$, as demonstrated in numerical simulations$^{157}$.

\bigskip

Figure 12 shows the outcome of an SPH simulation (performed using
GADGET-2$^{175}$) designed by Cuadra et al.$^{176}$ to follow the 
early dynamical evolution of a 
black hole binary inside
a circum-binary ring in Keplerian rotation and with initial profile
$\Sigma=\Sigma_0 (R_0/R)$.  The gap, already present as initial
condition, is maintained for the entire duration of the simulation
($\sim 1200$ dynamical times $ t_{\rm dyn}$) as gas piles up near 
$R\sim  3 a$ in response to the joint action
of the tidal torque offered by the binary and of the torques in the fluid. 
The binary soon starts to decay and to increase its eccentricity.
With time however, as illustrated in Figure 12, the ring, allowed to cool
on a few dynamical times, becomes unstable 
to non-axisymmetric perturbations that are transmitted through the entire structure.
The ring thus  spreads over a larger volume.  At this point the action of viscous torques becomes
progressively ineffective and the binary stalls.
Whether the binary continues to decay on a much longer timescale
is still matter of a debate, the process being reminiscent to
that of planet {\it migration} in circum-stellar disks$^{170-174}$.  

\bigskip 

\begin{figure}
\centering{
\resizebox{10cm}{!}{\includegraphics{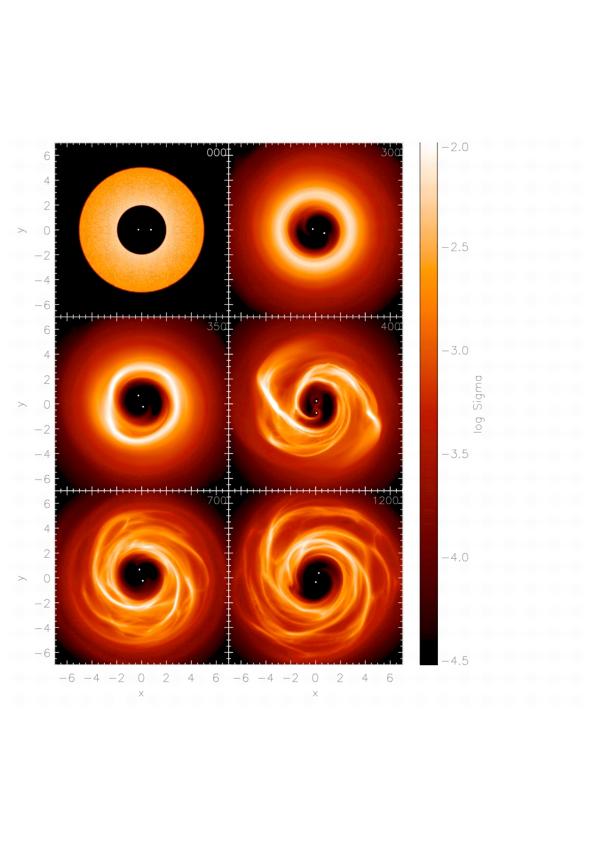}}
\caption{ Evolution of a black hole binary in a circum-binary
ring, as simulated by Cuadra et
al.$^{176}$.  The Figure shows the logarithmic maps of the disk column
density at different times, and the white dots indicate the current
position of the black holes.  The panel at $t = 0$ shows the smooth
initial condition. Until $t \sim 350\Omega_0^{-1}$ (where $\Omega_0$
is the Keplerian frequency of the binary at its initial orbital
separation $a_0$) material piles up at a distance comparable to $\sim
3 a_0$, as a result of the gravitational torques: the accretion disk
responds to the tidal field of the binary increasing its surface density at radii
nearest to the black holes. The
ring breaks due to its self-gravity, spreading its gas over the
original radial range. Stationary spiral patterns develop until the
simulation ends. The binary has mass ratio $q_\bh=0.3$ and the disk
mass is $M_{\rm disk}=0.2 M_{\bht}$. }  }
\label{fig:cuadra}
\end{figure}

Analytical and numerical studies of black hole migration in 
standard Keplerian accretion disks $^{174,177-181}$ have indicated
that migration is a slow process occurring on a timescale longer than the
viscous time $t_{\nu}$ at the inner edge of the 
disk where the gap is located ($ R_{\rm gap}\sim  2 a$).  If
$\Sigma_{\rm gap}$ is the unperturbed surface density of a reference
Keplerian disk at $R_{\rm gap}$ and $M_{\rm gap}=2\pi R^2_{\rm
gap}\Sigma_{\rm gap}$ the mass enclosed, the resulting migration time 
is crudely, $t_{\rm migration}\sim 
(M_\bht /M_{\rm gap}) t_{\nu}.$ 

\bigskip
Cuadra et al.$^{176}$ computed the migration time $t_{\rm migration}$ in a
self-regulated ($Q\sim 1$) circum-binary disk, starting from an estimate of $t_\nu$ 
based on a viscosity prescription that originates from self-gravity, and from
an appropriate estimate of the cooling time and of the opacity law 
of the gas in the disk.
Since $t_{\rm migration}$ is found to be 
proportional to $\Sigma_{\rm gap}^{-b}$ (with $b>0$), the maximum rate
of orbital decay can be inferred considering the highest $\Sigma_{\rm
gap}$ consistent with a Toomre parameter $Q\sim 1$. (Denser disks 
would likely fragment into stars rather than
remaining gaseous and this would destroy the hydrodynamical
interaction that leads to black hole binary migration.)  Following
Cuadra et al., binaries with mass $M_\bht<10^7\,\msun$, i.e.  binaries in the
frequency interval of {\it LISA} operation, are found to have migration times
$t_{\rm migration}\sim  10^{8-9}$ yr, while 
heavier binaries would instead stall, and/or decay via three-body encounters off 
stars$^{125}.$

\bigskip

If migration, as described in$^{176-179}$, continues down to smallest
scales, the circum-binary disk is no longer dominated by self-gravity
and takes the form of a Keplerian accretion disk.  In this regime, the
binary-disk interaction is better described in terms of a sequence of {\it
resonances}$^{181}$.  Azimuthal and longitudinal modes at the outer
Lindblad resonance are excited that promote the orbital decay of the
binary.  Whether black hole migration is
conducive to the domain in which gravitational waves take over and
guide the inspiral is still uncertain.  Should this occur, then below $a_\gw$ the
circum-binary disk would become no longer dynamically important.  Only
gravitational waves are expected to influence the orbital elements:
waves circularize the orbit and if $e$ is large at the end of black
hole migration, the two black holes may coalesce with still a residual
eccentricity which is imprinted in the gravitational wave
pattern$^{182}$.
\bigskip

The presence of the circum-binary accretion disk and of circum-stellar disks around
each black hole, unimportant at dynamical level during the GW phase,
can play a key role in determining whether the gravitational wave
signal from the binary is accompanied by a bright electromagnetic
counterpart.  This will be the subject of next Section.

\section{Preglow and Afterglow of coalescence events}

Consider a black hole binary crossing  the H $\to $ GW
boundary.  After a time $t_\gw$ it will then coalesce becoming
one of the loudest sources of gravitational waves observable in the
whole sky with {\it LISA}.  {\it LISA} will operate as an all--sky
monitor, detecting at the same time galactic and cosmological
sources. Given the difficulty to disentangle the different signals,
and the large number of noise sources, the accuracy in determining a
single source position is still matter of debate. A distinctive
electromagnetic counterpart to a {\it LISA} detection of a black
hole coalescence would have different payoffs, allowing to identify
the galaxy host, and to observe directly the
redshift of the source.  As a consequence, the simultaneous detection of a
gravitational wave and an electromagnetic  signal could be used to estimate
fundamental cosmological parameters as coalescing binaries are
standard {\it sirens}$^{183}$.  Detection of an electromagnetic 
counterpart would also improve our understanding of the accretion
physics of massive black holes, by comparing directly the 
luminosity with the black hole's masses and spins obtained directly
from the gravitational wave signal.  A large number of possible electromagnetic 
counterparts have been proposed in the last few years (see, e.g.,
Kocsis \& Loeb$^{184}$ and references therein).  In this review we will
discuss only few selected candidates, in relation  to the dynamical evolution
of the black holes during (the preglows) and after (the afterglows)
coalescence.

\subsection{Preglows}

As already discussed, the dynamical effect of a massive black hole
binary is to open a gap in its circum-binary accretion
disk. As the binary semi-major axis decays due to the interaction with
the disk itself, viscous dissipation maintains the outer 
edges of the gap at $R_{\rm gas}\sim 2 a$.
Only during the last phases of binary evolution, when
the semi-major axis decays due to gravitational wave emission, the
outer disk and binary evolution decouples.  The distance between the
black holes and the inner rim of the circum-binary disk (still moving
toward the center on the viscous timescale) increases. If an accretion
disk is present, in addition,  around the primary black hole (of an unequal mass
binary), the secondary creates the conditions for excavating an {\it
annular} gap, i.e. a hollow region between the circum-stellar disk around the
primary and the outer circum-binary disk.  As a consequence, the gas
orbiting at radii smaller than the outer edge of the inner 
circum-stellar disk is
driven inward under the action of gravitational torques by the
inspiraling secondary black hole.  Armitage \& Natarajan$^{185}$
showed that this effect amplifies the accretion rate onto the primary
that can largely exceed the Eddington limit just prior coalescence.
They suggested that a large fraction of the energy associated with
this extreme pre--coalescence accretion event is converted in thermal
energy of the gas accreting onto the primary. The increased
temperature triggers large outflows of gas, with velocities comparable
to the the orbital velocity of $\sim 10^4$ km s$^{-1}$. The authors
suggest that such strong high velocity outflows can be viewed as clear
signal of coalescing or recently coalesced black hole binaries.

\subsection{Afterglows}

Armitage \& Natarajan$^{185}$ showed that the evolution of the inner
edge of the gap causes a possibly observable preglow just before
coalescence.  In this section we discuss the consequences of the
evolution of the outer edge of the gap, resulting in an afterglow
observable years after coalescence.

\bigskip
When the black holes reach coalescence, the circum-binary disk is free
from gravitational torques and starts to spread inwards viscously
refilling the gap.  Milosavljevic \& Phinney$^{186}$
estimated that the timescale before complete refilling of the gap and
the turn--on of observable accretion activity is $\sim 7 \, (1+z)
(M/10^6 \msun)^{1.32}$ yr. They also constrained the luminosity and
spectrum of the turning--on AGN associated to the binary coalescence
event: $\sim 10^{43.5} (M/10^6 \msun)$ erg s$^{-1}$ are emitted
between 0.5--2 keV in the rest frame of the source and for hole's
masses in the {\it LISA} band. Based on models of the black hole
assembly$^{29,30,110}$, Dotti et al.$^{187}$ estimated that next
generation X--ray missions (such as {\it XEUS} or {\it
Constellation--X}) will be able to detect an X--ray afterglow for a
large fraction of the total coalescences detectable by {\it LISA} in
three years of operation ($\sim $ 9, 21, and 28 after 1, 5, and 20
yr from the GW detection, respectively).

\subsection{Kick-induced Shocks in circum-binary disks}

Black hole binaries recoil, at the time of their coalescence, due to
anisotropic emission of gravitational waves. The possibility that
recoiling black holes can produce an observable electromagnetic signature of a
recent coalescence, already suggested by Milosavljevic \&
Phinney$^{186}$, has been subject of a detailed discussion only
recently.

\bigskip
Lippai, Frei \& Haiman$^{188}$ simulated the response of a thin disk,
initially orbiting around a $10^6 \msun$ black hole, to kicks that
eject the black hole from the center. In particular they explored the
range $500 \,\kms < v_{\rm rec} < 4000\,\kms$, where ${\rm v_{rec}}$
is the kick velocity of the black hole recoiling either in the disk
plane or perpendicular to the disk.  The perturbation in the potential
well exerted by the black hole recoiling in the disk plane alters the
orbits of the particles forming the disk, resulting in the formation
of over-density regions distributed over a tightly wound spiral
caustic, as shown in the left panel of Figure 13. The
density enhancement within weeks after the ejection, can be more than
10 times larger than the unperturbed disk, and moves outward at the
speed ${\rm v}_{\rm rec}$. Density enhancements are observed also for
black holes recoiling perpendicularly to the disk plane (see
Figure 13 , right panel), but are much weaker ($\sim 
10 \%$ of the unperturbed value), and develops in caustic only after
$\gsim 1$ yr. Assuming that the gas forming the density enhancement is
shocked--heated, the authors estimated a prompt increase in the
emitted luminosity up to $L_{\rm rec}/L_{\rm Ed}\sim  6.3\times 10^{-4}
(1.6 \times 10^{-2})$, 70 days ($\sim  2$ yr) after the coalescence.
Lippai et al. found also that the converging velocity ($v_{\rm conv}$)
of the shocking regions is an increasing function of the radius, so
that the spectrum of the afterglow peaks at a characteristic photon
energy $k_B\, T_{\rm shock} \propto v_{\rm conv}^2 \sim 3$ eV 50 days
after coalescence, and grows up to $\sim 50 $ eV $\sim 2$ yr after.
\begin{figure} 
\centering{
\resizebox{15.cm}{!}{\includegraphics{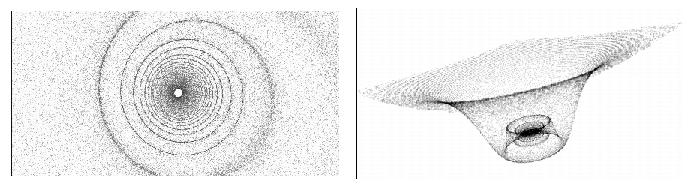}}
\caption{Output of two different runs presented in Lippai et
al.$^{188}$.  Left panel: face--on view of the perturbations triggered
in the accretion disk by a black hole recoiling in the disk plane. The
snapshot shows the output of a simulation with $v_{\rm rec}=500~\kms$,
90 days after the kick. Light shades correspond to regions of
unperturbed density, while dark shades to regions $\sim  10$ times
over-dense.  Right panel: areal view of a disk perturbed by a black
hole recoiling perpendicularly to the disk.  The snapshot is taken
after 1 week, and $v_{\rm rel}=500~\kms$.  In this case, the density
enhancement is noticeably weaker than the one observed in the coplanar
simulation, $\sim  10 \%$ of the initial value.} }
\label{fig:lippai}
\end{figure}
Schnittman \& Krolik$^{189}$ noticed that the disk considered in
Lippai et al.$^{188}$ is optically thick, and as a consequence the
radiation is absorbed and re--emitted with a thermal spectrum in the
infrared. They also applied their results to cosmological models of
structure formation, finding that $\sim 1-10$ of these afterglows
should be detectable in the {\it Spitzer, XMM-Newton}, and {\it
Chandra} surveys.

\subsection{Recoiling black holes in galaxy remnants} 

The presence of a recoiling black hole ejected from its nucleus can be
observed in galaxy merger remnants up to $10^8$ yr after the binary
coalescence. Different electromagnetic signatures have been proposed: the recoiling
black hole can be observed as an off-center QSO, if the hole itself is
able to retain a punctured disk at the coalescence$^{46}$. Volonteri
\& Madau$^{47}$  studied this process in a statistical context,
finding that tens of off-centered AGNs should be discovered by the
future {\it James Webb Space Telescope (JWST)}.  Episodic accretion
flares can be produced from the tidal disruption of stars during the
lifetime of the recoiling black hole$^{190,191}$.
\bigskip

\begin{figure} 
\label{fig:berni}
\centering{
\resizebox{11.cm}{!}{\includegraphics{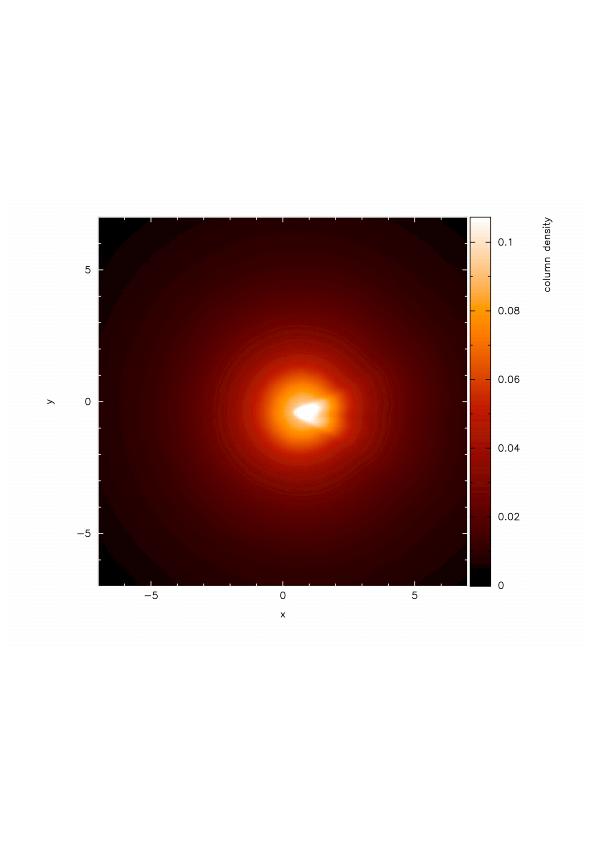}}
\caption{
Recoiling black hole in a massive elliptical galaxy
model$^{193}$.  The figure shows the color-coded
density map (in logarithmic scale) of the gas perturbed by the 
massive black hole on its return orbit, at its first pericentric passage.
The recoiling black hole (of $6 \times 10^9
\msun\sim 6\times 10^{-3} M_{\rm galaxy}$) is ejected with an initial $v_{\rm
rec}= 900$ km s$^{-1}$ (${\cal M}\sim 3$), and is moving on
the plane perpendicular to the line of sight. The box is 14 kpc aside. }}

\end{figure}

A recoiling massive black hole can also retain a ``hyper-compact stellar
system'' (HCSS) detectable in the optical as a luminous
cluster$^{190,192}$. The structural properties of HCSSs are defined by the
stellar distribution function of stars near the black holes prior
coalescence, and by the recoil velocity of the black hole remnant$^{192}$. The
authors estimate that the size and luminosity of HCSSs is similar to
that of a globular cluster, but can reach values typical of
ultra--compact dwarf galaxies if the host galaxies are very massive
and the kick velocities sufficiently small.  HCSSs are distinguishable
from globular clusters and dwarf galaxies because of the peculiar
velocity distribution of their stars: the internal velocities in a
HCSS are comparable to the recoil velocity, allowing for velocity
dispersions much larger than the ones observed in canonical globular
clusters.  The authors use cosmological models to constrain the number
of HCSSs observable in the local universe, finding that up to 100
HCSSs should be present in the Virgo cluster, even if only a few could
be bright enough to be confirmed as HCSSs by spectroscopical 
follow-ups.

\bigskip
A different signature of a recoiling black hole has been discussed in
Devecchi et al.$^{193}$.  The authors perform N-Body/SPH
simulations of black holes recoiling in massive elliptical galaxies, hosting a
hot, X--ray emitting, gas component. They discover that the
gravitational interaction between the moving black hole and the hot
medium excites a density perturbation in the gas, observable as
enhancement in the X--ray emission. If the black hole is moving with a
subsonic speed, a nearly spherical distribution of hot particles forms
around: this feature can live up to $\sim  10^8$ yr. If the black
hole is moving supersonically, it triggers an over-density in the form
of a Mach cone, as illustrated in Figure 14, observable for few tens of Myr.
The possibility to observe these features
(the spherical or conical X--ray enhancement) depends strongly on the
angular resolution of the X--ray detector.  
Thanks to the high spatial resolution, {\it Chandra} can detect recoiling black holes out to the distance of Virgo.

\section{Outstanding problems and conclusions}

The problem of formation and coalescence of dual black holes in
merging galaxies goes back to the influential paper of Begelman,
Blandford \& Rees$^{99}$.  Since then, our view on the formation of
cosmic structures and of their embedded black holes has improved
dramatically. We are now aware of the fact that dual, binary and
recoiling quasars are exclusive signposts of the process of galaxy
assembly. Their search in survey covering the entire electromagnetic
spectrum has then become compelling, as compelling is a deep
understanding of the dynamics of their formation, at theoretical
level.

 \bigskip
This review has touched key aspects of the long journey traveled by
the black holes on their path to coalescence, highlighting its
complexity.  The processes that are conducive, in merging galaxies, to
black hole coalescence are heterogeneous as the physical/dynamical
range that needs to be covered is enormous, going from the scale of a
cosmic merger ($\sim 100$ kpc) down to the tiny volume of a galaxy's
nuclear region of $\sim 10^{-3}$ pc. Only below this small scale,
gravitational waves alone drive the black hole to coalescence.  Given
the complexity of the problem, a step ahead has been carried out in
recent years thanks to major advances in numerical simulations.
Despite this effort there is no simulation at present that is capable
of following the pairing process in full realism, either in pure
stellar/gas-poor as well as in gas-rich environments, from
self-consistent cosmological conditions.

\bigskip
In stellar backgrounds there is still the need to explore in greater
detail loss cone refilling in the time-varying non-axisymmetric
gravitational potentials that form in the aftermath of a merger, and the
role of instabilities and of massive perturbers created in situ.  On a smaller scale,
when few body interactions become important, there is also the need to
include post-Newtonian black hole dynamics, as indicated in 
in a recent study$^{194}$.  In gas-rich
backgrounds, dissipation is known to play a key role in speeding black
hole inspiral, but often the thermodynamical description of the gas
has been treated in simple, idealized terms: the gas is multi-phase
and star formation, together with stellar/AGN feedback may play a
critical role in affecting the actual dynamics of the black holes.
The inclusion of these composite effects is a must for the future, and
we are working on these lines.  Low mass black holes at high
redshifts are likely to pair in gas-rich halos, while high mass black
holes at lower redshifts likely pair in gas-poor stellar backgrounds.
Thus exploring the dynamics in different habitat is of major
importance.

\bigskip

 The search of dual, binary and recoiling black holes, fed through
 accretion, will continue: imaging of galaxies in close interaction
 may unveil the presence of pairing black holes on kpc
 scales. Spectroscopically, the presence of large Doppler
 displacements in the emission line systems of quasars and
 AGNs$^{78,79,82}$ may reveal the presence of black holes in close
 orbital motion on scales close to a parsec.  A major advance
 will come from {\it LISA} in space and the Pulsar Timing Array
 experiment on Earth. Pulsar Timing Array will detect the emission of
 very massive black holes many year before coalescence ($< 10^6$
 yr), while LISA will detect the last year of inspiral and the final
 rapid coalescence. Thus, we have tools to trace the entire dynamical
 evolution of black holes using both the electromagnetic and
 gravitational wave windows.
 
\vskip 1 truecm

{\bf{{Acknowledgements}}}

\bigskip

We thank all the people that have had a long term collaboration with
the authors and contributed to many of the results discussed in this
review; Simone Callegari, Roberto Decarli, Bernadetta Devecchi, Fabio
Governato, Francesco Haardt, Stelios Kazantzidis, George Lake, Piero
Madau, Lucio Mayer, Carmen Montuori, Ben Moore, Tom Quinn, Alberto
Sesana, Joachim Stadel, Marta Volonteri, and James Wadsley.

\end{document}